\documentclass[fleqn,usenatbib]{mnras}

\usepackage{newtxtext,newtxmath}
\usepackage{subcaption}

\usepackage[T1]{fontenc}

\DeclareRobustCommand{\VAN}[3]{#2}
\let\VANthebibliography\thebibliography
\def\thebibliography{\DeclareRobustCommand{\VAN}[3]{##3}\VANthebibliography}

\usepackage{graphicx}	
\usepackage{amsmath}	
\usepackage{bm} 
\usepackage{multirow}

\usepackage{xeCJK}

\title[Misaligned circumbinary disc]{Misaligned circumbinary discs around unequal-mass eccentric binaries: alignment, morphology, and binary accretion variability}

\author[R. Yang et al.]{Ruiqi Yang (杨锐祺),$^{1}$\thanks{E-mail: yangrq9@mail2.sysu.edu.cn}
	Jeremy L. Smallwood ,$^{2}$\thanks{E-mail: smallj2@ou.edu}
	Hongping Deng (邓洪平),$^{3}$
	Ya-Ping Li (李亚平),$^{3}$
	\newauthor Alessia Franchini,$^{4}$
	Ruobing Dong (董若冰), $^{5,6}$
    Shang-Fei Liu (刘尚飞)$^{1,7}$
	\\
	$^{1}$School of Physics and Astronomy, Sun Yat-sen University, Zhuhai 519082, People's Republic of China\\
	$^{2}$Homer L. Dodge Department of Physics and Astronomy, University of Oklahoma, 440 W. Brooks St., Norman, OK 73019, USA\\
	$^{3}$Shanghai Astronomical Observatory, Chinese Academy of Sciences, Shanghai 200030, People's Republic of China\\
	$^{4}$Dipartimento di Fisica, Universit\`a degli Studi di Milano, Via Celoria 16, 20134, Milano (IT)\\
	$^{5}$Kavli Institute for Astronomy and Astrophysics, Peking University, Beijing 100871, People's Republic of China\\
	$^{6}$Department of Physics and Astronomy, University of Victoria, Victoria, BC, V8P 5C2, Canada\\
    $^{7}$CSST Science Center for the Guangdong-Hong Kong-Macau Greater Bay Area, Sun Yat-sen University, Zhuhai 519082, People's Republic of China
}

\date{Accepted XXX. Received YYY; in original form ZZZ}

\pubyear{2026}

\begin{document}
	\label{firstpage}
	\pagerange{\pageref{firstpage}--\pageref{lastpage}}
	\maketitle
	
	\begin{abstract}
    Binary systems are ubiquitous in the Universe and often host circumbinary discs that are misaligned with the binary orbital plane. Such misalignments can affect disc evolution and binary accretion variability. We here present 3D hydrodynamical simulations of circumbinary discs with initial tilts $i_0$ from $0^\circ$ to $180^\circ$, around eccentric binaries with secondary-to-primary mass ratios of $0.11-0.67$. We find that both the initial tilt and mass ratio can affect the long-term accretion variability in our simulations. Discs evolving towards polar and coplanar retrograde generally favour accretion onto the primary star, while discs evolving towards coplanar prograde generally favour accretion onto the secondary. We find preferential accretion ratio $\eta=\langle\dot{M_2}\rangle/\langle\dot{M_\mathrm{b}}\rangle$ to be a non-monotonic function of the mass ratio. For discs close to coplanar prograde alignment, $\eta$ increases with decreasing mass ratio, whereas for discs with $30^\circ \le i_0 \le 135^\circ$, $\eta$ decreases for smaller mass ratios. Polar discs show the lowest mass loss rates, slightly lower than those of coplanar prograde discs, while retrograde discs lose mass faster than their prograde counterparts. Discs that undergo strong warping or breaking experience rapid mass loss. Our findings provide insights into observed circumbinary discs and have implications for circumbinary planet formation.
	\end{abstract}
	
	
	
	\section{Introduction}
	Circumbinary discs play a significant role in various astrophysical systems, spanning protoplanetary discs around young stellar binaries to massive black hole binaries \citep{kley2012, barnes1998}. Understanding the interactions between binary systems and their surrounding discs has been a key focus of astrophysics for decades. It provides valuable insights into the evolution of these systems, including binary mass accretion rates, orbital dynamics, and observational signatures. In the context of stellar binaries, circumbinary discs are believed to form during the star formation phase \citep{boss1986,bonnell1994a,bonnell1994,kratter2008}, with observational evidence from systems such as DQ Tau, AK Sco, and UZ Tau E \citep{mathieu1997, czekala2016, alencar2003, anthonioz2015,czekala2015,simon2000,prato2002,jensen2007}. The evolution of these circumbinary discs might influence the formation and migration of circumbinary planets \citep{miranda2017,martin2019}.
    
    Numerous theoretical and numerical studies have investigated binary-disc interactions (see \citealt{lai2023a} for a detailed review). Binary resonant gravitational torques typically halt viscous inflow from the inner disc, creating a cavity a few binary separations wide \citep{artymowicz1994a,miranda2015}. Even around circular binaries, the cavity can become eccentric due to the binary torques, triggering the apsidal precession of the discs and potentially leading to long-term accretion variability \citep{munoz2016a,miranda2017,munoz2020a,ragusa2020}. Observationally, binary accretion variability can be characterised through both spectroscopic and photometric measurements. In coplanar prograde circumbinary discs, short-term accretion variability for eccentric binaries is modulated by the binary orbital period $P_\mathrm{b}$. For circular binaries with $q_\mathrm{b}=M_2/M_1\ge0.5$, the dominant modulation timescales of the variability are roughly 1 and 5 $P_\mathrm{b}$, driven by the density lumps that are periodically formed and destroyed at the inner edge of the disc. When $q_\mathrm{b}\le0.4$, the timescale changes into only 1 $P_\mathrm{b}$ \citep{macfadyen2008,farris2014a,shi2012,munoz2016a,miranda2017,moody2019,munoz2020a,franchini2023,cocchiararo2024,franchini2024}.
    
    Over longer timescales, equal-mass circular binaries tend to accrete at comparable rates onto each component. In contrast, eccentric binaries can exhibit preferential accretion, where one star accretes more than the other, with such preferential accretion alternating every few hundred orbits, possibly due to the apsidal precession of the eccentric inner disc \citep{dunhill2015, munoz2016a}. The TWA 3A system may be an example of this behaviour. With a binary eccentricity of 0.67, TWA 3A exhibits preferential accretion onto the primary star \citep{tofflemire2017b, tofflemire2019, anthonioz2015, czekala2021}, challenging earlier predictions that the secondary star would dominate accretion \citep{farris2014a, munoz2020a}. There are also many studies focused on modelling the coplanar retrograde circumbinary discs \citep{nixon2011,nixon2015,bankert2015,amaro-seoane2016, tiede2023}. A primary result from these studies is that the secondary typically dominates gas capture unless the binary mass ratio is close to unity or the capture radius of the secondary is very small.
    
    Most of the studies discussed above focused on coplanar circumbinary discs (prograde or retrograde). However, circumbinary discs can also be misaligned with respect to the binary orbital plane \citep{czekala2019}, potentially formed from chaotic turbulence in giant molecular clouds \citep{bate2003,bate2018}. Observationally, several misaligned circumbinary disc systems have been found. The pre-main-sequence binary KH 15D has a circumbinary disc tilted by about $3^\circ-15^\circ$ \citep{chiang2004,winn2004,capelo2012,smallwood2019,poon2021}. The circumbinary disc in system IRS 43 have a misalignment of about $60^\circ$ with the binary orbital plane \citep{brinch2016a}. HD 98800 B \citep{kennedy2019a} possesses a nearly polar gaseous circumbinary disc and 99 Herculis \citep{kennedy2012, smallwood2020} has a nearly polar circumbinary debris disc. A misaligned circumtriple disc has also been observed in the young hierarchical system GW Ori \citep{bi2020, kraus2020, smallwood2021,smallwood2025a}.
    
    In highly misaligned circumbinary discs around eccentric binaries, disc nodal precession tends to occur about the binary's eccentricity vector rather than its angular momentum vector. Due to the secular asymmetric potential from the binary, the disc undergoes tilt oscillations and eventually evolves into polar alignment due to dissipation within the disc \citep{verrier2009,farago2010, doolin2011, martin2017, lubow2018, zanazzi2018,smallwood2019}. This alignment alters the disc's morphology, with analytical studies suggesting smaller cavities in misaligned discs \citep{foucart2014,miranda2015} and numerical simulations linking binary-disc tilt to cavity size and torque distribution \citep{franchini2019, pirog2025}.
    
    The dynamics of circumbinary accretion in misaligned discs around unequal-mass eccentric binaries remains poorly explored. Recent research by \citet{smallwood2025} on equal-mass eccentric binaries identified preferential alternating accretion in coplanar discs and in discs that undergo breaking. However, the role that the binary mass ratio plays in misaligned systems remains unexplored. Early hydrodynamical star-formation simulations suggest that close binaries typically exhibit higher mass ratios than wider binaries \citep{bate2003}, and the mass ratio may markedly impact the binary accretion variability \citep{bate1997,bate2000,bate2002}. Given the variations in mass ratios and binary-disc tilts, the resonances in the disc become more complex \citep{miranda2015, miranda2017, martin2019}, making the evolution of these systems more intricate.
    
    \citet{pirog2025} modelled the misaligned circumbinary discs around unequal-mass circular binaries using a 2D grid-based code {\sc DISCO}. They decomposed the binary gravitational potential in the disc plane, focusing on 2D evolution of density and torque distribution. However, it is noteworthy that misaligned circumbinary discs tend to be associated with eccentric binaries, particularly those with high eccentricity \citep{dunhill2015,martin2019}. In this study, we expand the parameter space explored by \citet{smallwood2025} to investigate how mass ratio and binary-disc tilt influence disc alignment, morphology, and accretion variability.
    
    We here model misaligned circumbinary discs around eccentric binaries with varying mass ratios through 3D smoothed particle hydrodynamics (SPH) simulations. This approach enable us to capture three-dimensional effects, such as disc warping, breaking, and precession, which are absent by construction in \citet{pirog2025}. Our findings reveal that both mass ratio and binary-disc tilt significantly shape the evolution of the disc and the long-term accretion variability of the binary.
    
    This paper is organised as follows. Section~\ref{sec: methods} outlines the setup of our hydrodynamical simulations and the analysis methods employed. Section~\ref{sec: results} presents the key findings, including the evolution of disc alignment, disc morphology, binary accretion variability, and disc mass. In Section~\ref{sec: discussion}, we discuss the implications of our results, including disc breaking and potential numerical limitations. Finally, we summarise our conclusions in Section~\ref{sec: summary}.

	\section{Methods}
	\label{sec: methods}
	\subsection{Numerical Simulation Setup}
	We perform numerical simulations using the smoothed particle hydrodynamics (SPH) code {\sc phantom} \citep{price2018}, which demonstrates excellent agreement with the analytical framework describing tilted discs evolution of \citet{ogilvie1999}. This code has been extensively applied to study misaligned circumbinary discs \cite[e.g.,][]{nixon2015,martin2017,abod2022,smallwood2022}.
	
	We set up four eccentric binaries with mass ratios $q_\mathrm{b}=M_2/M_1\in\{0.67,\ 0.43,\ 0.25,\ 0.11\}$. The binary is modelled as two sink particles \citep{bate1995} with a total mass of $M_\mathrm{b} = M_1 + M_2$, where $M_1$ and $M_2$ denote the primary and secondary stars, respectively. Initially, the binary has a semi-major axis $a_\mathrm{b} = 1$ and eccentricity $e_\mathrm{b0}=0.5$. Although the binary orbit evolves over time due to feedback from the circumbinary disc, the disc mass in our simulations is too small to significantly alter the binary parameters \citep{martin2019}. Sink particles accrete gas particles that satisfy accretion checks, updating their mass and angular momentum \citep{price2018}. A gas particle is immediately accreted if it falls within 0.8 times the accretion radius. Otherwise, it is accreted if: its specific angular momentum is less than that of a Keplerian orbit at accretion radius, it is gravitationally bound to the sink particle, and it is more strongly bound to this sink than to any other. To reduce computational costs without significantly affecting the accretion rate, we set the accretion radii of the binary to $r_\mathrm{acc,1}=r_\mathrm{acc,2}=0.25a_\mathrm{b}$ \citep{smallwood2022}. Utilising Cartesian coordinates, the binary is initially positioned in the $x$-$y$ plane. The binary eccentricity vector aligns with the $x$-axis, and the binary angular momentum vector aligns with the $z$-axis. Table~\ref{tab:setup} summarises the main setups and outcomes of all the hydrodynamical simulations.
	
	Initially, the circumbinary disc is tilted relative to the $x$-$y$ plane by an angle $i_0$, varying from $0^\circ$ (prograde) to $180^\circ$ (retrograde) in increments of $15^\circ$. The disc is composed of $1\times10^6$ equal-mass SPH particles, with a total disc mass of $M_\mathrm{d}=10^{-3}M_\mathrm{b}$.	Although self-gravity can influence warps and tilt, the \citet{toomre1964} parameter $Q$ remains greater than 1 throughout all simulations, indicating that self-gravity effects are not important in this study. The inner radius of the disc is set to $r_\mathrm{in}=4a_\mathrm{b}$, and the outer radius to $r_\mathrm{out}=7a_\mathrm{b}$. For a coplanar disc, tidal truncation typically occurs at an inner radius of approximately 2-3$a_\mathrm{b}$ (\citealt{artymowicz1994a}). However, a tilted circumbinary disc, particularly a polar one, can extend closer to the binary orbit due to the weaker gravitational torque experienced by material off the orbital plane \citep{miranda2015, franchini2019}. The chosen value of $r_\mathrm{in}$ allows material to viscously drift inward during the initial stages before the disc reaches a quasi-steady state \citep{smallwood2022}, and it also mitigates artificially high accretion rates at the start of the simulation \citep{smallwood2021}. Disc breaking is in principle more likely in an extended disc. Following \citet{rabago2024}, we find that our choice of inner and outer radii prevents initial disc breaking, ensuring a stable early evolution while also supplying sufficient gas to the binary region.
	
	We adopt initial conditions similar to those in \citet{smallwood2025}. The gas surface density profile follows a power-law distribution, expressed by
	\begin{equation}
		\Sigma(r)=\Sigma_0\left(\frac{r}{r_\mathrm{in}}\right)^{-p},
		\label{eq:surface_density}
	\end{equation}
	where $\Sigma_0=1.54\times10^{-5}M_\mathrm{b}/a_\mathrm{b}^2$ is the initial density at the disc inner radius, $p=+3/2$ is the power law index, and $r$ is the spherical radius. We choose the locally isothermal equation of state according to \citet{farris2014a}, where the sound speed $c_\mathrm{s}$ is described as
	\begin{equation}
		c_\mathrm{s}=c_\mathrm{s0}\left(\frac{a_\mathrm{b}}{M_1+M_2}\right)^{q}\left(\frac{M_1}{r_1}+\frac{M_2}{r_2}\right)^{q}.
		\label{eq:sound_speed}
	\end{equation}
	Here $r_1$ and $r_2$ represent the distances from the primary and secondary stars, respectively, and $c_\mathrm{s0}\approx0.141$ is the sound speed at $r=a_\mathrm{b}$. This temperature profile ensures that, for $r_\mathrm{1,2} \ll r$, the disc behaves like an $\alpha$-disc near each star, while for $r_\mathrm{1,2} \gg r$, it resembles a circumbinary $\alpha$-disc.
	
	The disc pressure scale height is given by
	\begin{equation}
		H = c_\mathrm{s}/\Omega\propto r^{3/2-q},
		\label{eq:aspect_ratio}
	\end{equation}
	where $\Omega = (GM_\mathrm{b}/r^3)^{1/2}$ and $q = +3/4$. The initial disc aspect ratio at $r_\mathrm{in}$ is $H/r=0.1$. The kinematic viscosity $\nu$ is prescribed using \citet{shakura1973} viscosity $\alpha_\mathrm{SS}$
	\begin{equation}
		\nu=\alpha_\mathrm{SS}c_\mathrm{s} H.
	\end{equation}
	
	To model the $\alpha_\mathrm{SS}$ parameter in {\sc phantom}, we use the artificial viscosity $\alpha_\mathrm{AV}$ as described in \citet{lodato2010}
	\begin{equation}
		\alpha_\mathrm{SS}\approx\frac{\alpha_\mathrm{AV}}{10}\frac{\langle h\rangle}{H}
		\label{eq:viscosity}
	\end{equation}
	where the ratio $\langle h\rangle$ is the mean resolution length at a given radius, $H$ is the disc scale height, and the $\alpha_\mathrm{SS}$ parameter is set to 0.01 in our simulations. The circumbinary disc is initially resolved with average smoothing length per scale height $\langle h\rangle/H \approx 0.1386$, and $\langle h\rangle/H$ can vary azimuthally due to the warps in the disc \citep{fairbairn2021, deng2022}.

    \begin{table*}
    \caption{Summary of simulation parameters and outcomes. The first column represents the simulation name. The second column describes the binary mass ratio ($q_\mathrm{b}$). The third column denotes the initial binary-disc tilt ($i_0$). The fourth column lists the accretion radius in units of the binary separation ($r_\mathrm{acc}$). The fifth column shows the disc alignment, where C indicates coplanar alignment, P denotes polar alignment, and the asterisk superscript (*) marks systems exhibiting disc breaking. The sixth column gives the simulation duration in binary orbital periods ($t$). The seventh column shows the estimated tilt decay timescale in binary orbital periods ($\tau_\mathrm{dec}$). The last column indicates the accretion type, where P indicates the primary dominates the accretion, S denotes the secondary dominates the accretion, and A marks the alternating preferential accretion.}
    \label{tab:setup}
    \centering
    \begin{minipage}{0.495\linewidth}
    \centering
    \resizebox{\linewidth}{!}{
        \begin{tabular}{lccccccc}
            \hline
            \hline
            Name & $q_\mathrm{b}$ & $i_0$ $[^\circ]$ & $r_\mathrm{acc}$ $[a_\mathrm{b}]$ & Align & $t$ [$P_\mathrm{b}$] & $\tau_\mathrm{dec}$ [$P_\mathrm{b}$] & Acc. type \\
            \hline
            0p67i0    & 0.67 & 0    & 0.25 & C      & 5000 & -     & A \\
            0p67i15   & 0.67 & 15   & 0.25 & C      & 5000 & 25884 & A \\
            0p67i30   & 0.67 & 30   & 0.25 & C      & 5000 & 32541 & A \\
            0p67i45   & 0.67 & 45   & 0.25 & P$^*$  & 5000 & -     & A $\rightarrow$ P \\
            0p67i60   & 0.67 & 60   & 0.25 & P      & 5000 & 4078  & P \\
            0p67i75   & 0.67 & 75   & 0.25 & P      & 5000 & 4014  & P \\
            0p67i90   & 0.67 & 90   & 0.25 & P      & 5000 & -     & P \\
            0p67i105  & 0.67 & 105  & 0.25 & P      & 5000 & 4538  & P \\
            0p67i120  & 0.67 & 120  & 0.25 & P      & 5000 & 3747  & P \\
            0p67i135  & 0.67 & 135  & 0.25 & P$^*$  & 5000 & -     & A $\rightarrow$ P \\
            0p67i150  & 0.67 & 150  & 0.25 & C      & 5000 & 47283 & P \\
            0p67i165  & 0.67 & 165  & 0.25 & C      & 5000 & 12331 & S \\
            0p67i180  & 0.67 & 180  & 0.25 & C      & 5000 & - & S \\
            s0p67i0  & 0.67 & 0  & 0.025 & C      & 2000 & - & S \\
            s0p67i90  & 0.67 & 90  & 0.025 & P      & 1000 & - & P \\
            s0p67i150  & 0.67 & 150  & 0.025 & C      & 2000 & - & P \\
            s0p67i165  & 0.67 & 165  & 0.025 & C      & 2000 & - & P \\
            s0p67i180  & 0.67 & 180  & 0.025 & C      & 2000 & - & S \\
            \hline
            0p43i0    & 0.43 & 0    & 0.25 & C      & 5000 & - & S \\
            0p43i15   & 0.43 & 15   & 0.25 & C      & 5000 & 28873 & S \\
            0p43i30   & 0.43 & 30   & 0.25 & C      & 5000 & 37201 & A \\
            0p43i45   & 0.43 & 45   & 0.25 & P$^*$  & 5000 & - & A $\rightarrow$ P \\
            0p43i60   & 0.43 & 60   & 0.25 & P      & 5000 & 7015 & P \\
            0p43i75   & 0.43 & 75   & 0.25 & P      & 5000 & 6202 & P \\
            0p43i90   & 0.43 & 90   & 0.25 & P      & 5000 & - & P \\
            0p43i105  & 0.43 & 105  & 0.25 & P      & 5000 & 6214 & P \\
            0p43i120  & 0.43 & 120  & 0.25 & P      & 5000 & 6186 & P \\
            0p43i135  & 0.43 & 135  & 0.25 & P$^*$  & 5000 & - & P \\
            0p43i150  & 0.43 & 150  & 0.25 & P$^*$  & 5000 & - & P \\
            0p43i165  & 0.43 & 165  & 0.25 & C      & 5000 & 16792 & P \\
            0p43i180  & 0.43 & 180  & 0.25 & C      & 5000 & - & P \\
            s0p43i150  & 0.43 & 150  & 0.025 & C      & 500 & - & P \\
            s0p43i165  & 0.43 & 165  & 0.025 & C      & 2000 & - & P \\
            s0p43i180  & 0.43 & 180  & 0.025 & C      & 2000 & - & S \\
            \hline
        \end{tabular}
    }
    \end{minipage}%
    \hfill
    \begin{minipage}{0.495\linewidth}
    \centering
    \resizebox{\linewidth}{!}{
        \begin{tabular}{lccccccc}
            \hline
            \hline
            Name & $q_\mathrm{b}$ & $i_0$ $[^\circ]$ & $r_\mathrm{acc}$ $[a_\mathrm{b}]$ & Align & $t$ [$P_\mathrm{b}$] & $\tau_\mathrm{dec}$ [$P_\mathrm{b}$] & Acc. type \\
            \hline
            0p25i0    & 0.25 & 0    & 0.25 & C      & 5000 & - & S \\
            0p25i15   & 0.25 & 15   & 0.25 & C      & 5000 & 42613 & S \\
            0p25i30   & 0.25 & 30   & 0.25 & C      & 5000 & 63948 & S \\
            0p25i45   & 0.25 & 45   & 0.25 & P      & 5000 & - & P \\
            0p25i60   & 0.25 & 60   & 0.25 & P      & 5000 & - & P \\
            0p25i75   & 0.25 & 75   & 0.25 & P      & 5000 & - & P \\
            0p25i90   & 0.25 & 90   & 0.25 & P      & 5000 & - & P \\
            0p25i105  & 0.25 & 105  & 0.25 & P      & 5000 & - & P \\
            0p25i120  & 0.25 & 120  & 0.25 & P      & 5000 & - & P \\
            0p25i135  & 0.25 & 135  & 0.25 & P      & 5000 & - & P \\
            0p25i150  & 0.25 & 150  & 0.25 & P$^*$  & 5000 & - & P \\
            0p25i165  & 0.25 & 165  & 0.25 & C      & 5000 & 23431 & P \\
            0p25i180  & 0.25 & 180  & 0.25 & C      & 5000 & - & P \\
            s0p25i150  & 0.25 & 150  & 0.025 & C      & 1000 & - & P \\
            s0p25i165  & 0.25 & 165  & 0.025 & C      & 2000 & - & P \\
            s0p25i180  & 0.25 & 180  & 0.025 & C      & 2000 & - & P \\
            \hline
            0p11i0   & 0.11 & 0    & 0.25 & C      & 5000 & - & S \\
            0p11i15  & 0.11 & 15   & 0.25 & C      & 5000 & - & S \\
            0p11i30  & 0.11 & 30   & 0.25 & C      & 5000 & - & P \\
            0p11i45  & 0.11 & 45   & 0.25 & P      & 5000 & - & P \\
            0p11i60  & 0.11 & 60   & 0.25 & P      & 5000 & - & P \\
            0p11i75  & 0.11 & 75   & 0.25 & P      & 5000 & - & P \\
            0p11i90  & 0.11 & 90   & 0.25 & P      & 5000 & - & P \\
            0p11i105 & 0.11 & 105  & 0.25 & P      & 5000 & - & P \\
            0p11i120 & 0.11 & 120  & 0.25 & P      & 5000 & - & P \\
            0p11i135 & 0.11 & 135  & 0.25 & P      & 5000 & - & P \\
            0p11i150 & 0.11 & 150  & 0.25 & C      & 5000 & - & P \\
            0p11i165 & 0.11 & 165  & 0.25 & C      & 5000 & - & P \\
            0p11i180 & 0.11 & 180  & 0.25 & C      & 5000 & - & P \\
            s0p11i0  & 0.11 & 0  & 0.025 & C      & 2000 & - & S \\
            s0p11i90  & 0.11 & 90  & 0.025 & P      & 1000 & - & P \\
            s0p11i150  & 0.11 & 150  & 0.025 & C      & 1000 & - & P \\
            s0p11i165  & 0.11 & 165  & 0.025 & C      & 500 & - & P \\
            s0p11i180  & 0.11 & 180  & 0.025 & C      & 2000 & - & S \\
            \hline
        \end{tabular}
    }
    \end{minipage}
\end{table*}
    
	\subsection{Analysis Procedure}
	For disc analysis, we divide the disc uniformly into 300 bins in spherical radius, spanning from $1a_\mathrm{b}$ to $30a_\mathrm{b}$ to accommodate the disc's viscous expansion over time. Within each bin, we calculate azimuthally averaged quantities such as surface density, tilt, longitude of the ascending node, eccentricity, and warp amplitude. 
    
	We compute the angular momentum of the binary as
	\begin{equation}
		J_\mathrm{b}=\mu\sqrt{G(M_1+M_2)a_\mathrm{b}(1-e_\mathrm{b}^2)},
	\end{equation}
	where $\mu=(M_1M_2)/(M_1+M_2)$ is the reduced mass of the binary. The angular momentum of the disc is given by
	\begin{equation}
		J_\mathrm{d}=\int_{r_\mathrm{in}}^{r_\mathrm{out}}2\pi r^3 \Sigma(r)\Omega\ \mathrm{d}r.
	\end{equation}

	We calculate the tilt between an annulus in the disc at radius $r$ and the binary's angular momentum by
	\begin{equation}
		i_\mathrm{bd}(r) = \cos^{-1} (\bm{\hat{l}}_\mathrm{b}\cdot \bm{\hat{l}}_\mathrm{d}(r)),
		\label{eq:ibd}
	\end{equation}
	where $\bm{\hat{l}}_\mathrm{b}$ is the unit vector in the direction of the binary angular momentum and $\bm{\hat{l}}_\mathrm{d}(r)$ is the unit vector in the direction of the annulus angular momentum vector. As the disc undergoes dynamic warping and potential breaking, we introduce the density-weighted average inclination to capture its overall structure
	\begin{equation}
		i=\frac{\int_{r_\mathrm{in}}^{r_\mathrm{out}}\ 2\pi r \Sigma(r)i_\mathrm{bd}(r)\ \mathrm{d}r}{M_\mathrm{tot}},
	\end{equation}
	where $\Sigma(r)$ denotes the surface density at radius $r$, $i_\mathrm{bd}(r)$ depicts the inclination of the annulus at radius $r$, and $M_\mathrm{tot}$ refers to the instantaneous total disc mass. The annulus' longitude of the ascending node at radius $r$ is defined as 
	\begin{equation}
		\phi(r) = \tan^{-1}\left(\frac{\bm{\hat{l}}_\mathrm{d}(r)\cdot(\bm{\hat{l}}_\mathrm{b}\times \bm{\hat{e}}_\mathrm{b})}{\bm{\hat{l}}_\mathrm{d}(r)\cdot\bm{\hat{e}}_\mathrm{b}}\right)+\frac{\pi}{2}.
	\end{equation}
	The density-weighted average of the longitude of the ascending node is given by
	\begin{equation}
		\phi = \frac{\int_{r_\mathrm{in}}^{r_\mathrm{out}}\ 2\pi r \Sigma(r)\phi(r)\ \mathrm{d}r}{M_\mathrm{tot}}.
	\end{equation}
	We also define the eccentricity of the annulus at radius $r$ as
	\begin{equation}
		e(r) = \sqrt{\bm{\hat{e}}_\mathrm{d}(r)\cdot\bm{\hat{e}}_\mathrm{d}(r)},
	\end{equation}
	where $\bm{\hat{e}}_\mathrm{d}(r)$ is the unit vector in the direction of the annulus eccentricity vector. The density-weighted average of the disc eccentricity is
	\begin{equation}
		e = \frac{\int_{r_\mathrm{in}}^{r_\mathrm{out}}\ 2\pi r \Sigma(r)e(r)\ \mathrm{d}r}{M_\mathrm{tot}}.
	\end{equation}
	The disc warp amplitude, characterised by the dimensionless parameter $\psi$, which is defined as \citep{pringle1992,ogilvie1999}:
	\begin{equation}
		\psi=r\left|\frac{\partial \bm{\hat{l}}_\mathrm{d}(r)}{\partial r}\right|.
	\end{equation}

	For binary accretion analysis, we compute the long-term binary accretion rates by averaging over $(25/2\pi)\,P_\mathrm{b}$, where $P_\mathrm{b}$ is the initial binary period. As accretion is primarily influenced by the inner disc structure, we also examine the eccentricity of the cavity. To account for varying cavity sizes with different binary-disc tilts, we adopt $r_\mathrm{cav} = 2a_\mathrm{b}$ for discs evolving toward polar alignment and $r_\mathrm{cav} = 3a_\mathrm{b}$ for those evolving toward coplanar alignment (see section~\ref{subsubsec:cavity}; also \citealt{miranda2015,franchini2019}).

	\section{Results}
	\label{sec: results}
	\subsection{Disc alignment}
	\label{subsec: disc alignment}

    \begin{figure}
		\centering
            \includegraphics[width=\linewidth]{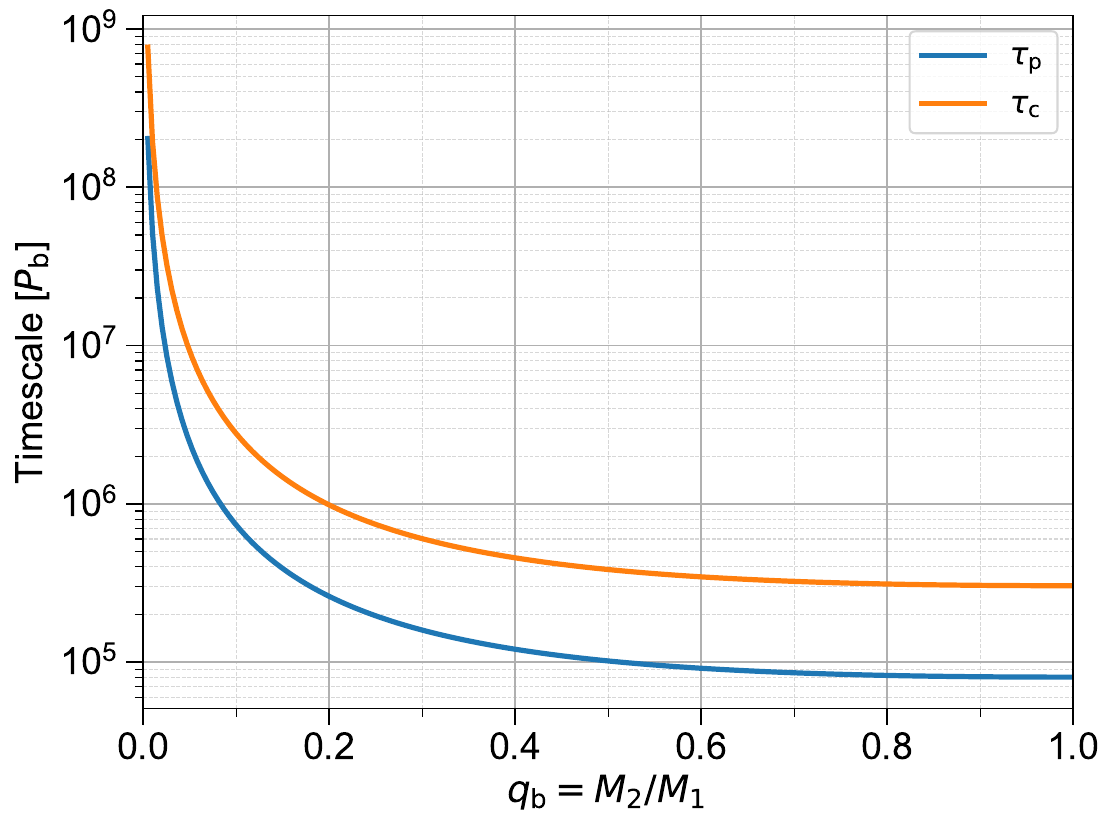}
		\caption{Analytically estimated timescales for coplanar and polar alignments as a function of binary mass ratio $q_\mathrm{b}$. The blue (orange) curve shows the analytically calculated polar (coplanar) alignment timescale, $\tau_\mathrm{p}$ ($\tau_\mathrm{c}$). The timescales for both alignments increase monotonically with smaller mass ratios, and the polar alignment consistently occurs faster than coplanar alignment.}
		\label{fig:tau}
	\end{figure}
    
	\begin{figure*}
		\centering
		\begin{subfigure}[b]{0.48\textwidth}
			\centering
			\includegraphics[width=\textwidth]{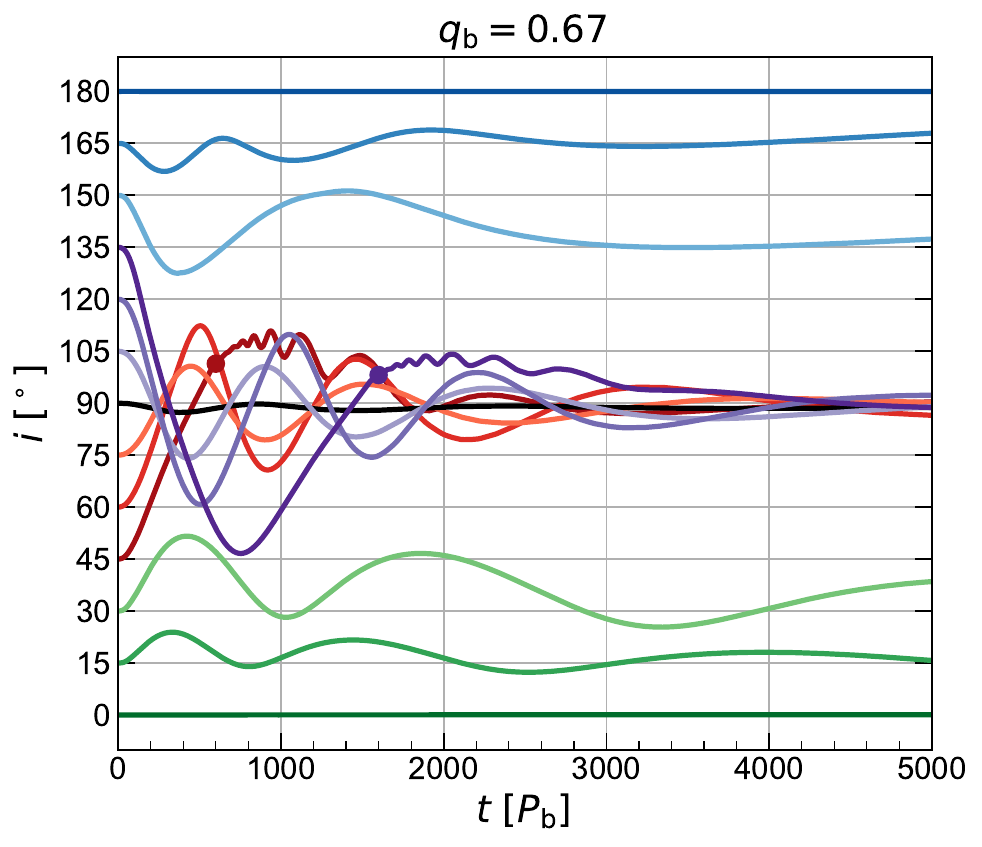}
			\label{fig:inc_evo_0p67}
		\end{subfigure}
		\begin{subfigure}[b]{0.48\textwidth}
			\centering
			\includegraphics[width=\textwidth]{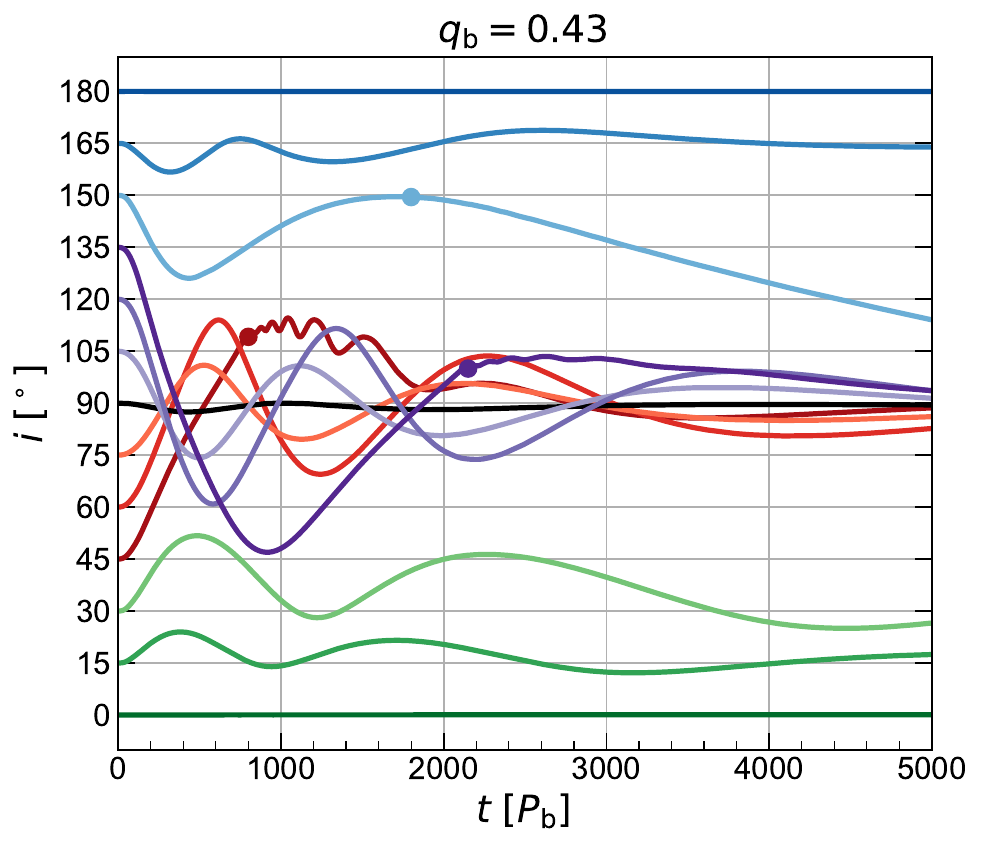}
			\label{fig:inc_evo_0p43}
		\end{subfigure}
		\\
		\begin{subfigure}[b]{0.48\textwidth}
			\centering
			\includegraphics[width=\textwidth]{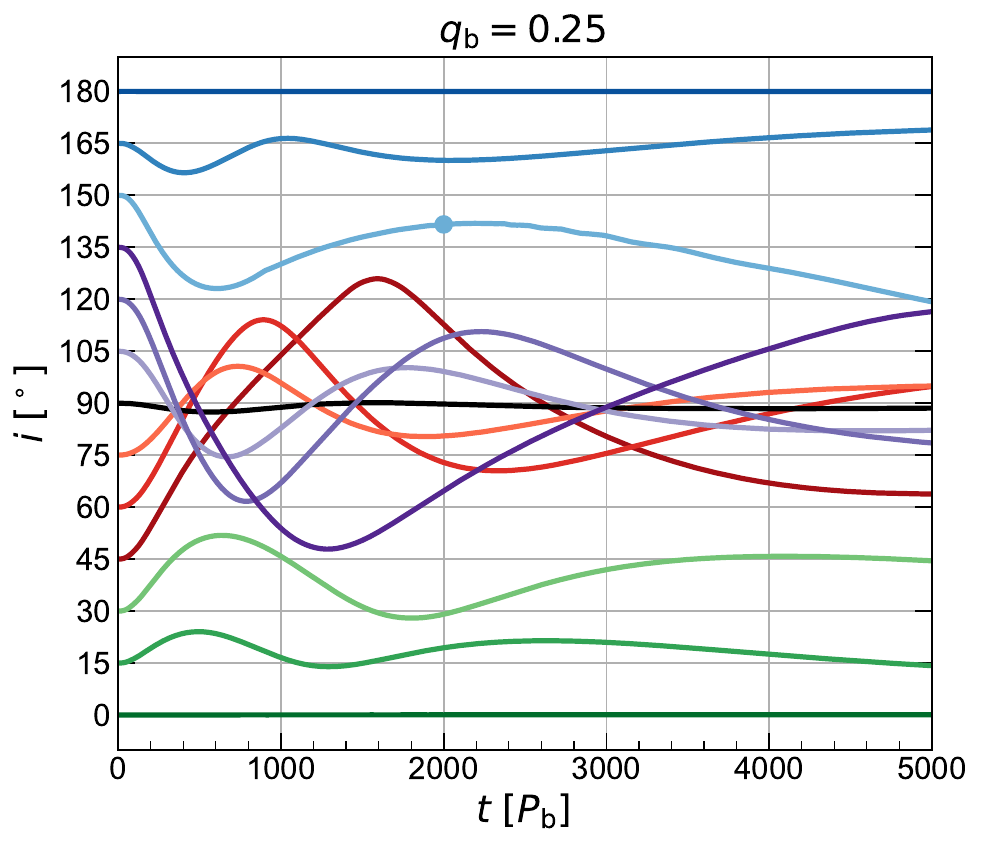}
			\label{fig:inc_evo_0p25}
		\end{subfigure}
		\begin{subfigure}[b]{0.48\textwidth}
			\centering
			\includegraphics[width=\textwidth]{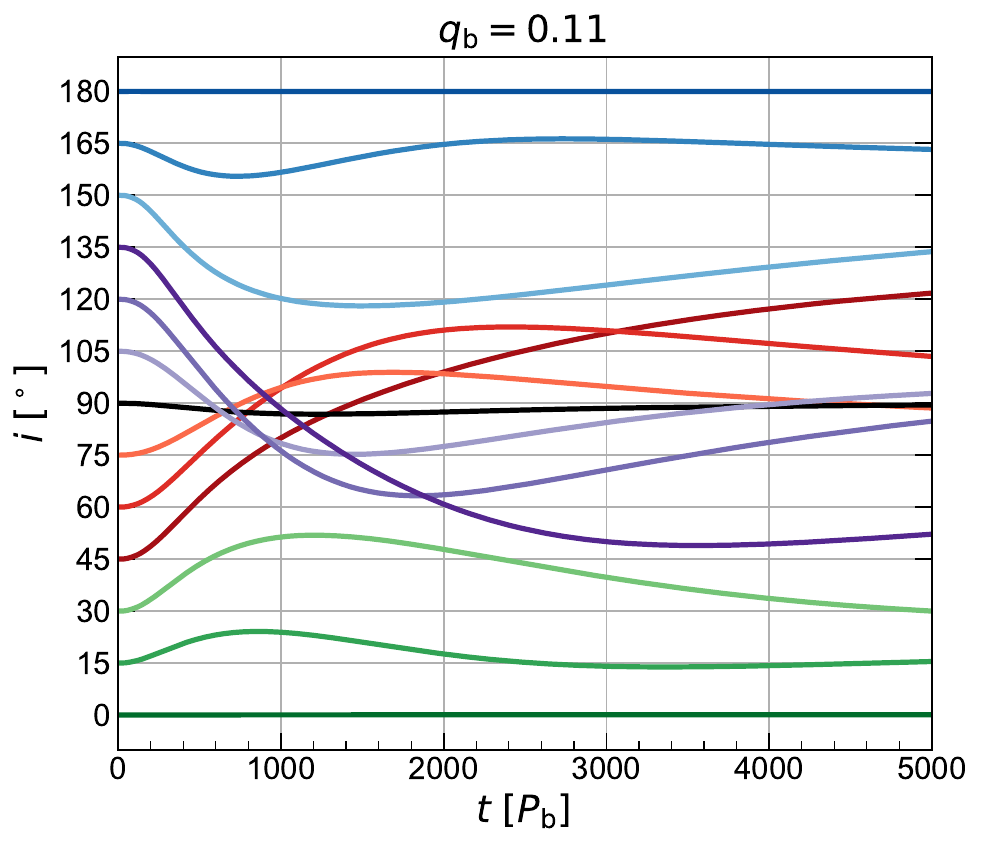}
			\label{fig:inc_evo_0p11}
		\end{subfigure}
		\caption{The evolution of binary-disc tilt for four binary mass ratios: $q_\mathrm{b}=0.67$ (top left), $q_\mathrm{b}=0.43$ (top right), $q_\mathrm{b}=0.25$ (bottom left) and $q_\mathrm{b}=0.11$ (bottom right), in units of initial binary orbital period $P_\mathrm{b}$. Different colours represent various initial binary-disc tilts ($i_0$): greens and blues indicate discs evolving towards coplanar prograde and retrograde alignments, respectively; reds and purples represent discs evolving towards polar alignment; black denotes discs that are initially polar. Coloured circles denote the disc breaking times. The tilts oscillate, with oscillation amplitudes gradually damping over time. As the binary mass ratio becomes smaller, the oscillation timescale noticeably increases.}
		\label{fig:inc_evo}
	\end{figure*}
	
		\begin{figure*}
		\centering
		\begin{subfigure}[b]{0.48\textwidth}
			\centering
			\includegraphics[width=\textwidth]{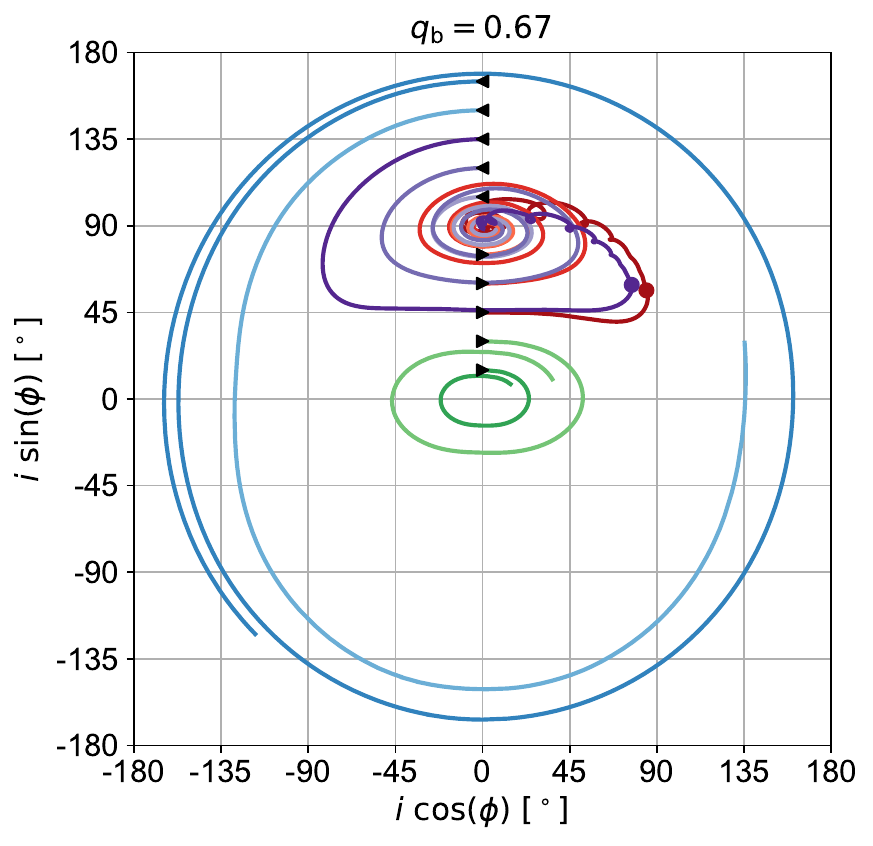}
			\label{fig:phase_0p67}
		\end{subfigure}
		\begin{subfigure}[b]{0.48\textwidth}
			\centering
			\includegraphics[width=\textwidth]{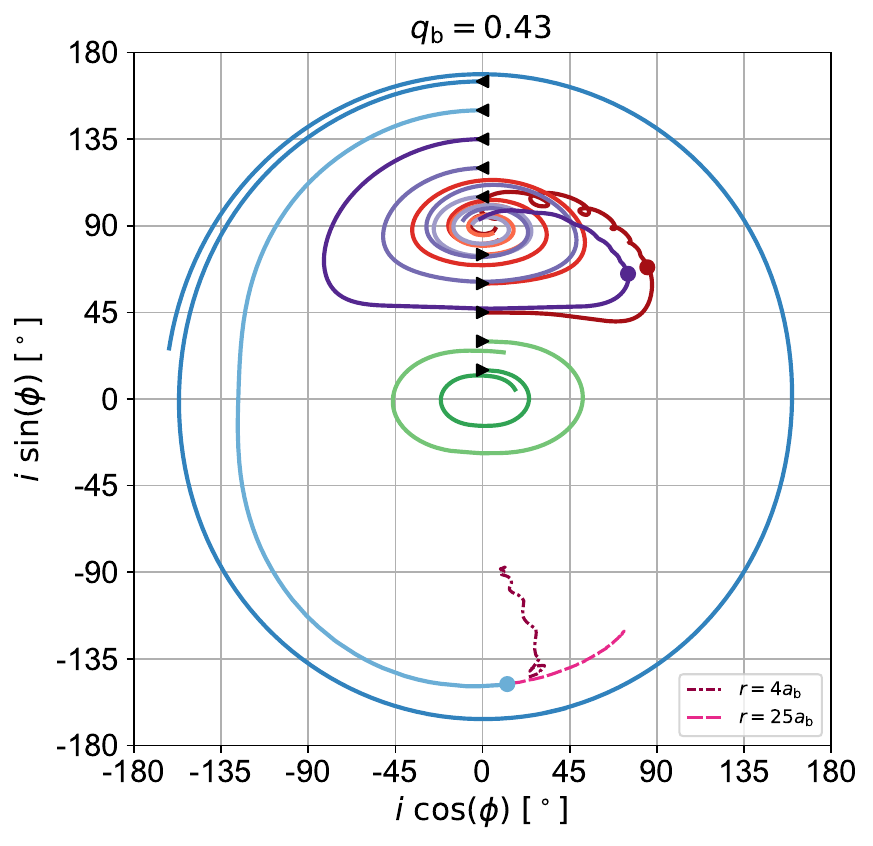}
			\label{fig:phase_0p43}
		\end{subfigure}
		\\
		\begin{subfigure}[b]{0.48\textwidth}
			\centering
			\includegraphics[width=\textwidth]{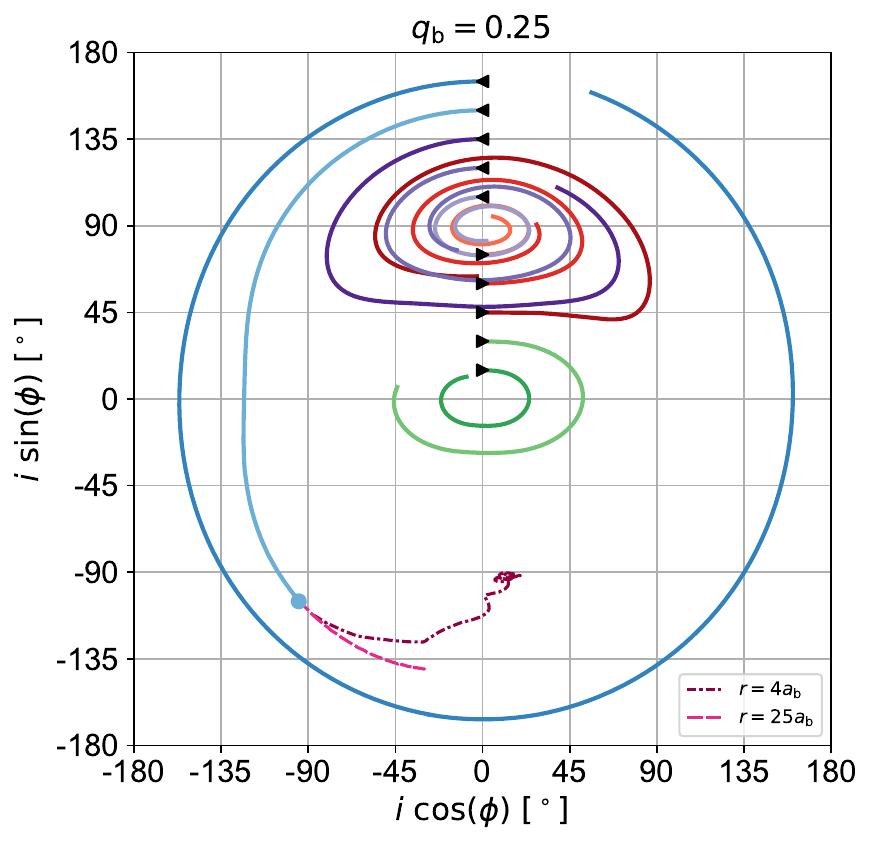}
			\label{fig:phase_0p25}
		\end{subfigure}
		\begin{subfigure}[b]{0.48\textwidth}
			\centering
			\includegraphics[width=\textwidth]{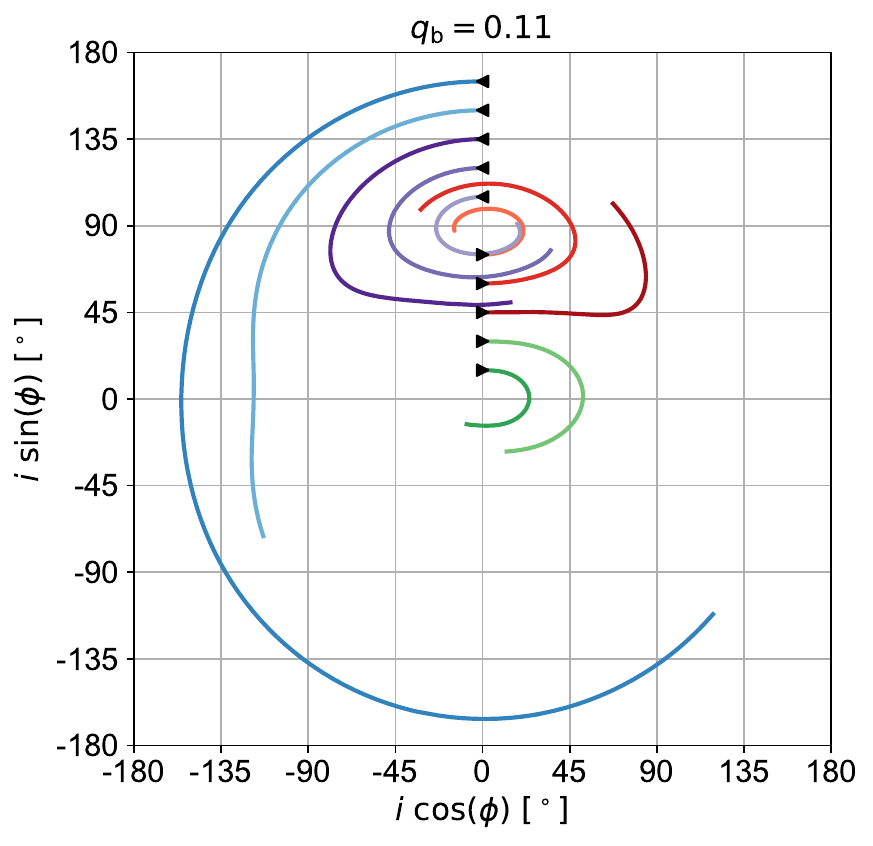}
			\label{fig:phase_0p11}
		\end{subfigure}
		\caption{The $i\cos{\phi}-i\sin{\phi}$ phase spaces for four binary mass ratios: $q_\mathrm{b}=0.67$ (top left), $q_\mathrm{b}=0.43$ (top right), $q_\mathrm{b}=0.25$ (bottom left) and $q_\mathrm{b}=0.11$ (bottom right). Different coloured curves correspond to various initial binary-disc tilts, consistent with the definition in Fig.~\ref{fig:inc_evo}. Black arrows indicate the precession direction of the discs. Coloured circles denote the disc breaking times. The maroon dash-dotted curve represents inner disc ($r=4 a_\mathrm{b}$), while the magenta dashed curve represents outer disc ($r=25 a_\mathrm{b}$). Note that results for $i_0=0^\circ$, $90^\circ$, and $180^\circ$ are not included, as the longitude of the ascending node is undefined for coplanar discs and precession is negligible for initially polar discs.}
		\label{fig:phase_evo}
	\end{figure*}

	A misaligned low-mass circumbinary disc around a circular binary undergoes nodal precession about the binary's angular momentum vector while maintaining a constant tilt. This behaviour is called \textit{circulation} \citep{nixon2012, facchini2013, lodato2013, foucart2013}. However, the dynamics differs in the case of an eccentric binary system. When the tilt between the circumbinary disc and the binary orbital plane exceeds a critical value, the disc's precession axis shifts from the binary's angular momentum vector to the binary's eccentricity vector. This distinct behaviour is instead called \textit{libration} \citep{verrier2009, farago2010, doolin2011, aly2015}.
	
	From \citet{martin2019}, the analytic tilt criterion was derived to determine whether a circumbinary disc will evolve towards coplanar alignment or polar alignment. In our setup, the ratio of disc-to-binary angular momentum, $j=J_\mathrm{d}/J_\mathrm{b}$, is sufficiently small, which corresponds to the lower $j$-branch defined in \citet{martin2019}. The analytic tilt criterion is given by:
	\begin{equation}
		\cos i_\mathrm{min}=\frac{\sqrt{5}e_\mathrm{b0}\sqrt{4e_\mathrm{b0}^2-4j_0^2(1-e_\mathrm{b0}^2)+1}-2j_0(1-e_\mathrm{b0}^2)}{1+4e_\mathrm{b0}^2},
	\end{equation}
	where $e_\mathrm{b0}$ is the initial binary eccentricity, and the term $j_0=J_\mathrm{d0}/J_\mathrm{b0}$ is the ratio of the initial disc-to-binary angular momentum. In our simulations,  $e_\mathrm{b0}=0.5$ and $j_0\approx$ 0.0110, 0.0125, 0.0165, 0.0293 for $q_\mathrm{b}=$ 0.67, 0.43, 0.25, 0.11, respectively. These values yield the lower and upper tilt criteria $i_\mathrm{lower}\approx$ $38.5^\circ$, $38.6^\circ$, $38.9^\circ$, $39.8^\circ$ and $i_\mathrm{upper}\approx$ $141.5^\circ$, $141.4^\circ$, $141.1^\circ$, $140.2^\circ$, respectively. Discs with $i_0\le i_\mathrm{lower}$ will evolve towards a coplanar prograde alignment. Discs with $i_\mathrm{lower} \le i_0\le i_\mathrm{upper}$ will evolve towards a polar alignment. Discs with $i_0 \ge i_\mathrm{upper}$ will evolve towards a coplanar retrograde alignment.
    
    The eccentric binary generates a secular non-axisymmetrical potential with respect to its angular momentum vector, causing the disc tilt to oscillate as it precesses. Viscous dissipation of the warp induced in the disc brings it to either alignment or polar alignment \citep{martin2017, lubow2018, zanazzi2018,martin2018}. The timescale of disc alignment is inversely proportional to the disc viscosity $\alpha_\mathrm{SS}$ \citep{lubow2018, smallwood2019}, which is given by
	\begin{equation}
		\tau = \frac{1}{\alpha_\mathrm{SS}}\left(\frac{H}{r}\right)^{2}\frac{\Omega_\mathrm{b}}{\Omega^2_\mathrm{d}}.
		\label{eq:tau}
	\end{equation}
	The disc nodal precession frequency $\Omega_\mathrm{d}$ is described as 
	\begin{equation}
		\Omega_\mathrm{d}=k\left\langle\left(\frac{a_\mathrm{b}}{r}\right)^{7/2}\right\rangle\Omega_\mathrm{b},
		\label{eq:Omega_d}
	\end{equation}
	where $k$ is an integrated coefficient dependent on binary eccentricity $e_\mathrm{b}$ and mass ratio $q_\mathrm{b}$, which is given by
	\begin{equation}
		k = \begin{cases}
			-\frac34\sqrt{1+3e_\mathrm{b}^2-4e_\mathrm{b}^4}\frac{M_1M_2}{M_\mathrm{b}^2}, & \text{for coplanar alignment}. \\
			\frac{3\sqrt{5}}{4}e_\mathrm{b}\sqrt{1+4e_\mathrm{b}^2}\frac{M_1M_2}{M_\mathrm{b}^2}, & \text{for polar alignment}.
		\end{cases}
		\label{eq:k}
	\end{equation}
	The bracketed term from equation~\ref{eq:Omega_d} is defined as
	\begin{equation}
		\left\langle\left(\frac{a_\mathrm{b}}{r}\right)^{7/2}\right\rangle = \frac{\int_{r_\mathrm{in}}^{r_\mathrm{out}}\ \Sigma r^3\Omega(a_\mathrm{b}/r)^{7/2}\ \mathrm{d}r}{\int_{r_\mathrm{in}}^{r_\mathrm{out}}\ \Sigma r^3\Omega\ \mathrm{d}r},
        \label{eq:bracket}
	\end{equation}
	where $\Omega=\sqrt{G(M_1+M_2)/r^3}$ is the Keplerian angular frequency at radius $r$. Note that if the disc breaks, the two broken discs will precess at different rates and equations above become invalid.

    To make an analytical prediction, we assume fixed parameters from the initial conditions for both coplanar and polar discs: surface density ($\Sigma$), viscosity parameter ($\alpha_\mathrm{SS}$), aspect ratio ($H/r$), outer disc radius $r_\mathrm{out}$, and binary eccentricity ($e_\mathrm{b}$). The only parameter adjusted is the inner disc radius, which varies between coplanar and polar alignments due to reduced binary torque effects in misaligned discs. The weakened binary torques in polar discs create smaller truncation radii compared to coplanar discs \citep{artymowicz1994a,miranda2015,hirsh2020}. Accordingly, we adopt $r_\mathrm{in,p}=1.5a_\mathrm{b}$ and $r_\mathrm{in,c}=3a_\mathrm{b}$ as our fiducial inner disc radii for polar and coplanar discs, respectively. Using equations~\ref{eq:tau} to~\ref{eq:bracket}, we calculate the alignment timescales for coplanar and polar discs as a function of mass ratio, as shown in Figure~\ref{fig:tau}. It shows that the alignment timescales for both configurations increase monotonically with smaller mass ratios, and the polar alignment consistently occurs faster than coplanar alignment.
    
    Figure~\ref{fig:inc_evo} illustrates the evolution of the binary-disc tilt $i$ for four mass ratios in our simulations. Coloured circles denote the disc breaking times. Note that there are two tilt criteria, $i_\mathrm{lower}\sim39^\circ$ and $i_\mathrm{upper}\sim141^\circ$. Discs with $i_\mathrm{lower}\le i_0 \le i_\mathrm{upper}$ evolve towards polar alignments (red, black and purple curves). Discs with $i_0 < i_\mathrm{lower}$ or $i_0 > i_\mathrm{upper}$ evolve towards coplanar prograde (green curves) or coplanar retrograde (blue curves). All simulations show tilt oscillation, and the amplitude of oscillation gradually dampens during alignment, with nearly polar discs achieving final alignment faster than nearly coplanar discs. To estimate the tilt decay timescales in our simulations, we adopt equation $\tau_\mathrm{dec} = (t_2-t_1)/\log{[(i_1-i_\mathrm{s})/(i_2-i_\mathrm{s})]}$ \citep{martin2018,smallwood2020}. Here, $i_1$ and $i_2$ are the first two local maxima (minima) of the inclination for initially prograde (retrograde) discs and their corresponding times $t_1$ and $t_2$, respectively. The stationary inclination $i_\mathrm{s}$ is calculated from the initial conditions \citep{martin2019}. $i_\mathrm{s}=0^\circ$, $89.3^\circ$, and $180^\circ$ for discs evolving towards coplanar prograde, polar, and coplanar retrograde alignments, respectively. In some simulations, the discs become strongly warped or broken, or exhibit only a single local extremum in inclination. In these cases, the tilt decay timescale cannot be reliably estimated. In addition, we attempt to compute alignment timescales by requiring that the density-weighted average of the longitude of the ascending node of the disc and the azimuthal angle of the binary eccentricity vector differ by less than $1^\circ$ \citep{smallwood2020}. However, none of our simulations satisfy this criterion, and we therefore do not calculate the estimated alignment timescales. The tilt decay timescales that can be estimated are listed in Table~\ref{tab:setup}. The tilt decay timescale increases with decreasing binary mass ratio, a trend that is consistent with our analytical calculations.
    
    Figure~\ref{fig:phase_evo} shows the $i\cos{\phi}-i\sin{\phi}$ phase spaces for four binary mass ratios. The distance from the origin represents the binary-disc tilt $i$. The azimuthal position corresponds to the ascending node of the circumbinary disc $\phi$. The black arrow directs the disc precession direction. Coloured circles denote the disc breaking times. The prograde disc precesses clockwise, whereas the retrograde disc precesses anticlockwise. This is due to different torque direction exerted by the binary onto the disc. The discs with $i_0<i_\mathrm{lower}$ and $i_0>i_\mathrm{upper}$ undergo circulation and precess about the direction and the counter direction of binary angular momentum vector, respectively. The discs with $i_\mathrm{lower}\le i_0\le i_\mathrm{upper}$ precess about the upper librating region, which is located at $i\approx90^\circ$ and $\phi\approx 90^\circ$. For a smaller binary mass ratio, the nodal precession timescale increases, leading to a longer evolutionary timescale in the direction of the longitude of the ascending node.
    
    For $q_\mathrm{b}=0.43$ and $q_\mathrm{b}=0.25$, discs with $i_0=150^\circ$ finally precess about the lower librating region, $i\approx90^\circ$ and $\phi\approx-90^\circ$. Both discs undergo breaking. To highlight the differences between the inner and outer disc regions during disc breaking, we employ equation~\ref{eq:ibd} to define the instantaneous tilt of an annulus in the disc at a characteristic radius: the maroon dash-dotted curve traces the inner disc evolution at $r=4 a_\mathrm{b}$, and the magenta dashed curve traces outer disc evolution at $r=25 a_\mathrm{b}$. Given that truncation radii of the inner disc range from $2a_\mathrm{b}$ to $4a_\mathrm{b}$ for discs with $i_0=150^\circ$, $r = 4a_\mathrm{b}$ can effectively represent the inner disc region. Disc breaks in the inner region and quickly forms a inner polar disc, whereas the outer disc maintains its inclination. As the break propagates outward, the density of the inner polar disc gradually increases, causing the density-weighted average inclination of the entire disc to evolve toward about $90^\circ$ (see Fig.~\ref{fig:inc_evo}).  We do not present results for different disc regions in the broken discs with $i_0 = 45^\circ$ and $135^\circ$ because, by the time disc breaks, the entire disc has already aligned to nearly polar. Thus, there is no significant difference in tilt between the inner and outer regions.
    
    \subsection{Disc morphology}
    \label{subsec:morphology}
    
    As discussed in section~\ref{subsec: disc alignment}, circumbinary discs that are initially inclined within the tilt criteria will evolve towards polar alignment and undergo tilt oscillation. During this alignment process, the disc can experience strong warping or even breaking, which significantly alter its morphology. We analyse the surface density of discs with varying instantaneous tilts, aiming to identify correlations between the initial tilt and the resulting disc morphology, which will ultimately impact the accretion process.
    
    \subsubsection{Cavity size}
    \label{subsubsec:cavity}
    \begin{figure}
        \centering
            \includegraphics[width=0.99\linewidth]{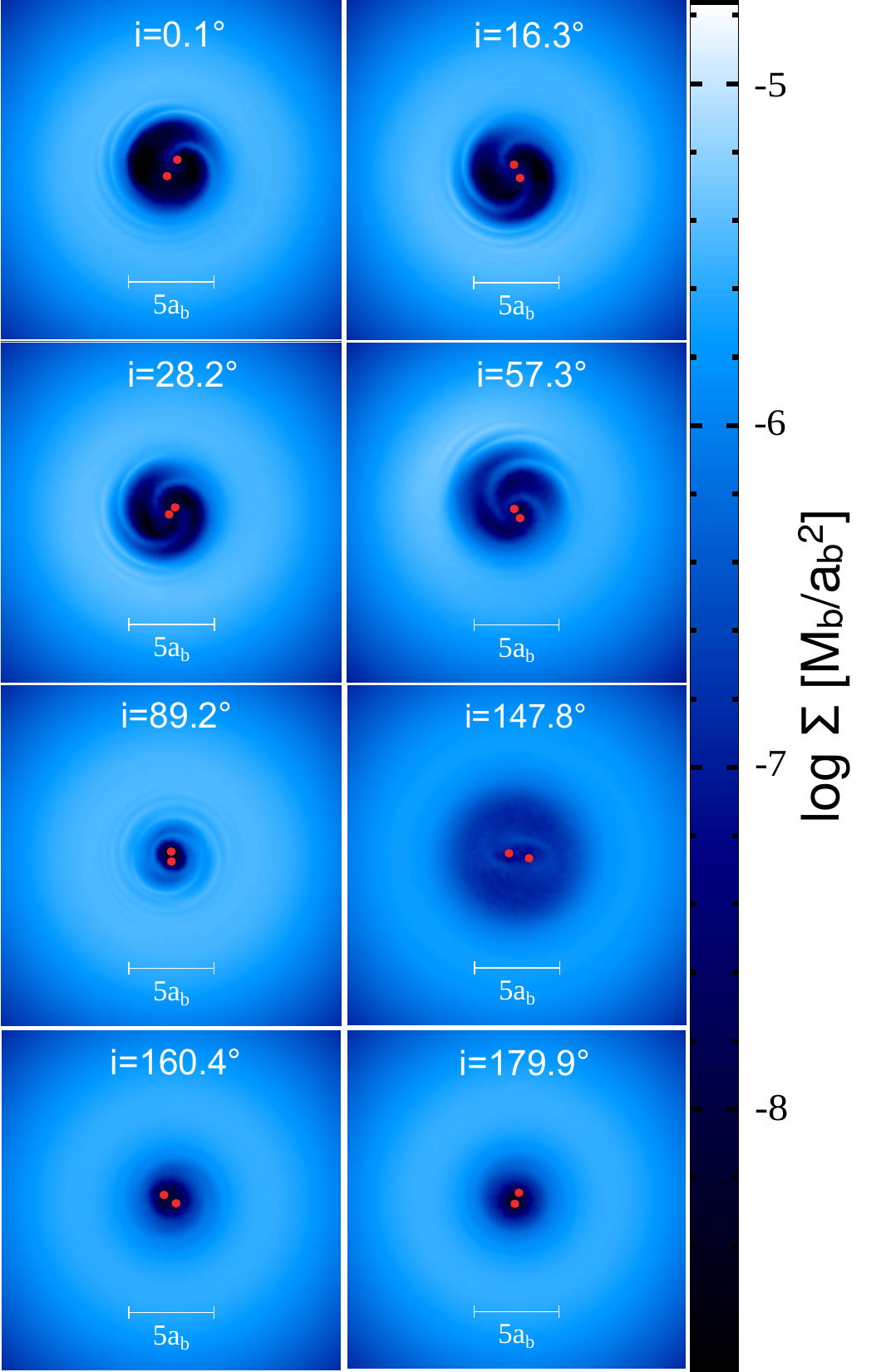}
            \caption{Surface density snapshots of circumbinary discs with different instantaneous binary-disc tilts $i$ at 1000 binary orbital periods, viewed opposite to the direction of the disc's angular momentum. From top left to bottom right, $i$ are: $0.1^\circ$ (0p67i0), $16.3^\circ$ (0p67i15), $28.2^\circ$ (0p67i30), $57.3^\circ$ (0p67i135), $89.2^\circ$ (0p67i90), $147.8^\circ$ (0p67i150), $160.4^\circ$ (0p67i165), $179.9^\circ$ (0p67i180). The two red markers indicate positions of the binary components, shown in their instantaneous orbital phase.}
        \label{fig:surface}
    \end{figure}
	\begin{figure}
		\centering
		\includegraphics[width=\linewidth]{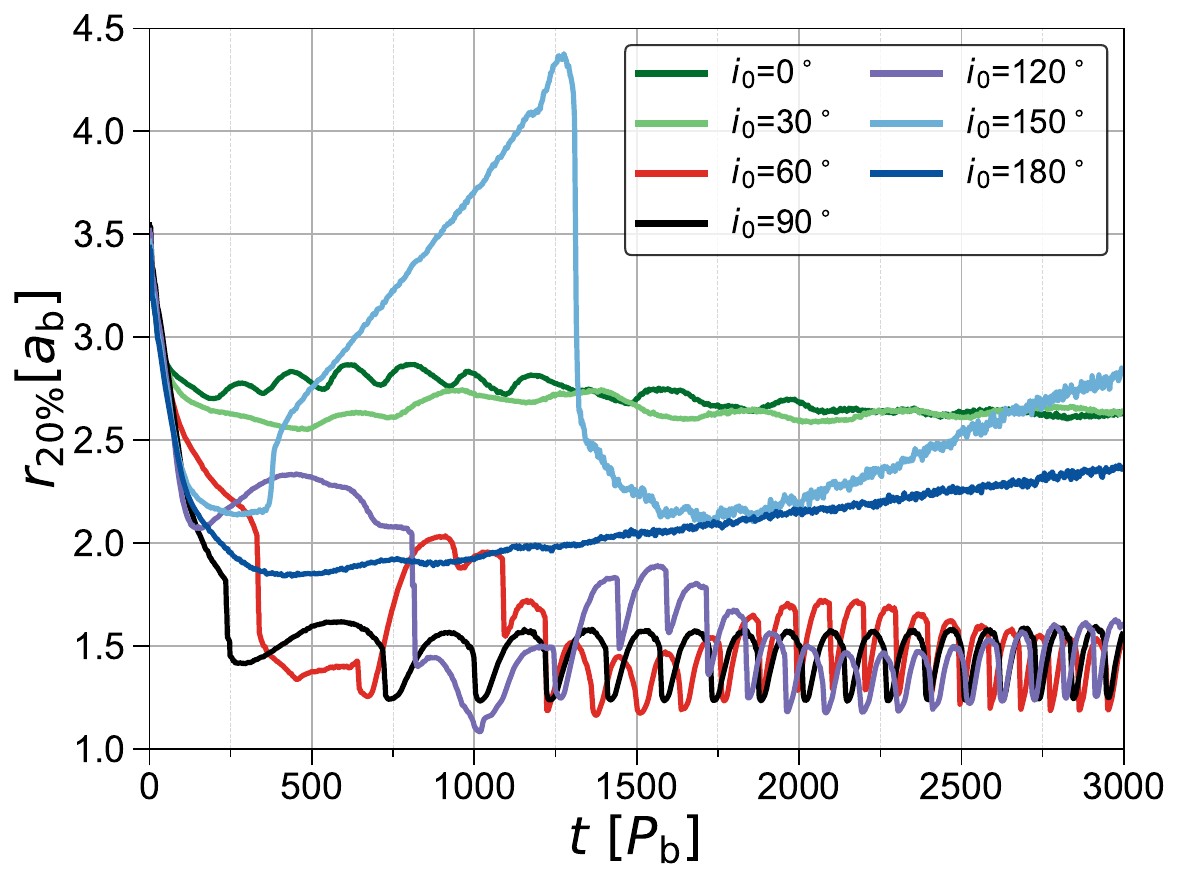}
		\caption{The evolution of reference radius $r_{20\%}$, as a function of binary orbital periods ($P_\mathrm{b}$). $r_{20\%}$ is the reference radius where the surface density reaches its 20\% of the maximum. Different curve colours represent simulations with varying initial binary-disc tilts for a binary mass ratio of $q_\mathrm{b} = 0.67$.}
		\label{fig:trunc_r}
	\end{figure}
    
    Figure~\ref{fig:surface} shows surface density snapshots of circumbinary discs with different instantaneous binary-disc tilts $i$: $0.1^\circ$ (0p67i0), $16.3^\circ$ (0p67i15), $28.2^\circ$ (0p67i30), $57.3^\circ$ (0p67i135), $89.2^\circ$ (0p67i90), $147.8^\circ$ (0p67i150), $160.4^\circ$ (0p67i165), $179.9^\circ$ (0p67i180) at 1000 $P_\mathrm{b}$. These snapshots are viewed opposite to the direction of the disc's angular momentum, with the binary components marked as red points. By this stage, the surface density in the inner regions of the discs has reached a steady state.
	
    For discs evolving towards coplanar prograde alignment, we see multiple spiral density waves. Unlike discs around circular binaries, those orbiting around eccentric binaries lack asymmetric high density lumps in the innermost regions \citep{miranda2017}. This difference arises because the gravitational forces in eccentric binary systems exhibit pattern frequencies both larger and lower than the binary orbital frequency, hindering the formation of such sensitive lump structures.
    
    The central cavity size decreases significantly as the tilt increases from coplanar prograde to polar. This trend is consistent with the theory of circumbinary disc resonances: as the disc tilts, the torque exerted by the binary on the disc weakens, leading to a smaller cavity. Such phenomenon suggests that the cavity size might serve as a valuable diagnostic for binary-disc misalignment in future observations \cite[e.g.,][]{franchini2019}.
	
	In retrograde discs, spiral density waves are also present but are notably weaker than those in prograde discs. This is consistent with prior studies of retrograde circumbinary discs \citep{nixon2011, nixon2015, tiede2023}. These studies have shown that resonances can be generated between an eccentric binary system and a retrograde disc. The components of the eccentric binary potential that rotate in the opposite direction to the binary, rotate in the same direction to the disc. When the binary eccentricity is sufficiently high, these components exert torques strong enough to excite spiral waves and carve out a central cavity. The coplanar prograde disc maintains a more extended cavity structure than its retrograde counterpart. However, the system with $i=147^\circ$ (0p67i150) exhibits pronounced warping, inducing an extended low-density region that surpasses even the prograde cavity size. This behaviour suggests that extreme warping can temporarily enlarge cavities, though subsequent alignment processes will reduce the cavity size as the disc evolves toward stable coplanar retrograde alignment.
	
	 We define the reference radius $r_{20\%}$ to quantitatively characterise the inner disc cavity size as the radius where the surface density reaches 20\% of its maximum value. The alternative reference radius $r_{10\%}$, used in earlier studies \citep{miranda2017, pirog2025}, is avoided here because, in highly misaligned discs, the cavity shrinks to the point where $r_{10\%}$ falls within the binary orbit, making it ill-defined in our simulations.

    \begin{figure}
	\centering
        \includegraphics[width=\linewidth]{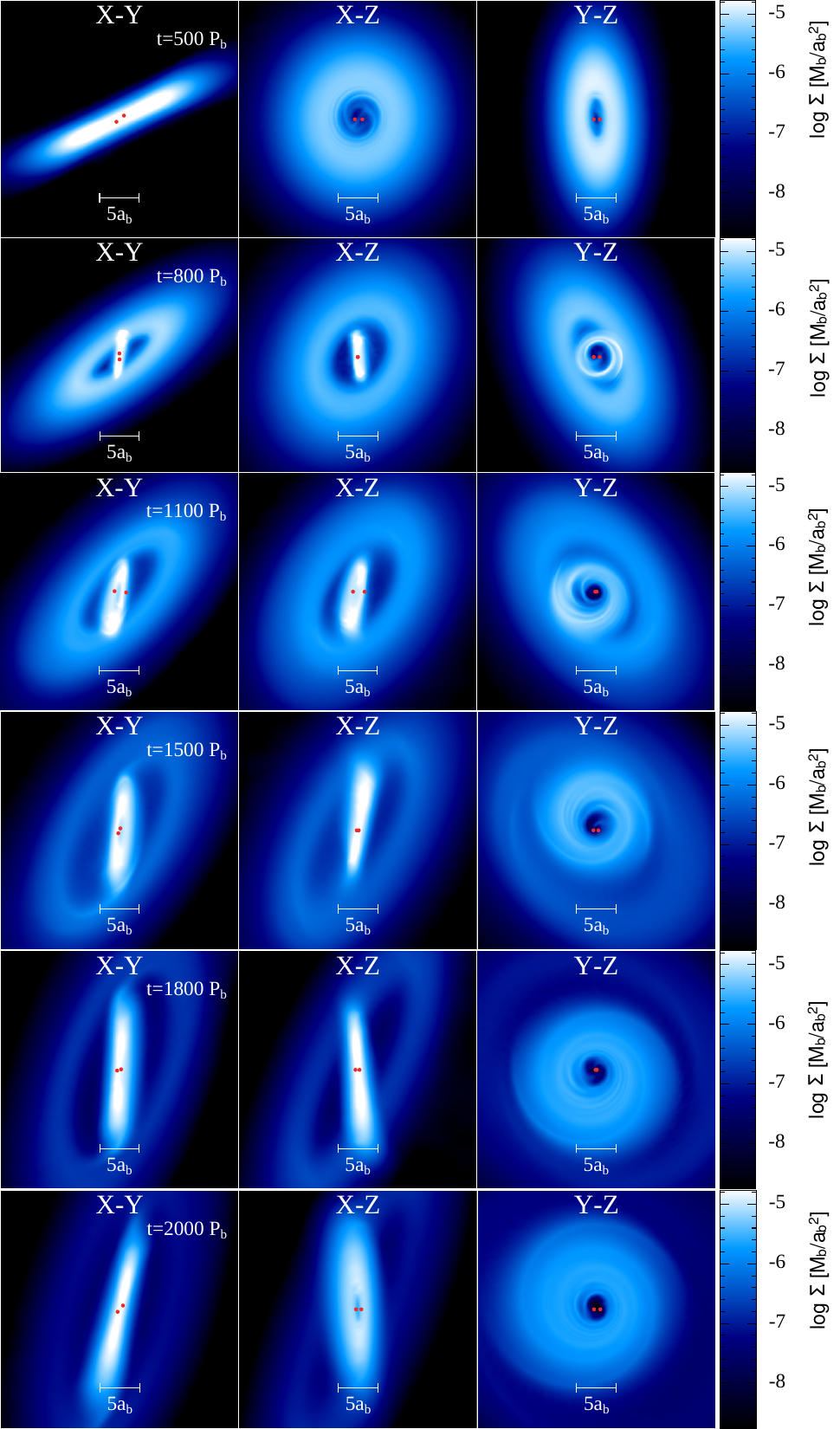}
        \caption{Surface density snapshots of a circumbinary disc with initial binary-disc tilt $i_0 = 45^\circ$ and binary mass ratio $q_\mathrm{b}=0.67$ during breaking. The sub-panels shows the $x$-$y$ plane (left), $x-z$ plane (middle) and $y-z$ plane (right). Snapshots are selected ranging from $t = 500 P_\mathrm{b}$ to $t = 2000 P_\mathrm{b}$.}
	\label{fig:dens_evo_0p67i45}
	\end{figure}
	
    \begin{figure}
        \centering
        \begin{subfigure}[b]{\linewidth}
            \centering
            \includegraphics[width=\textwidth]{Fig/Break_evo/break_0p67_45.pdf}
            \label{fig:break_evo_0p67}
        \end{subfigure}
        \caption{The evolution of the surface density $\Sigma$, amplitude of the warp $\psi$, eccentricity $e$, binary-annulus tilt $i_\mathrm{bd}$, and longitude of the ascending node $\phi$, as a function of radius $r$, for the case of $i_0 = 45^\circ$ in $q_\mathrm{b}=0.67$. Different colours, from dark red to ocean blue, denote different time, from $t=500P_\mathrm{b}$ to $1200P_\mathrm{b}$, with intervals of $100P_\mathrm{b}$. With the disc break propagating outward, the inner disc pumps its eccentricity, precesses about the binary eccentricity vector.}
	\label{fig:break_evo}
	\end{figure}
    
	Figure~\ref{fig:trunc_r} presents the evolution of $r_{20\%}$ for discs with $i_0\in\{0^\circ,\ 30^\circ,\ 60^\circ,\ 90^\circ,\ 120^\circ,\ 150^\circ,\ 180^\circ\}$ and $q_\mathrm{b}=0.67$. In the coplanar prograde disc, $r_{20\%}$ exceeds that of polar and coplanar retrograde discs due to the stronger gravitational torque from the binary. It stabilises at approximately $2.5-3.0a_\mathrm{b}$, consistent with previous analytical and numerical findings \citep{artymowicz1994a, miranda2015, hirsh2020}. For discs evolving towards polar alignment, $r_{20\%}$ oscillates around 1.5$a_\mathrm{b}$, aligning with predictions from \citet{miranda2015}. The oscillation of $r_{20\%}$ is caused by the discrepancy between the output interval (based on the initial binary period) and the actual binary period, which evolves over time. This accumulated difference reaches one binary orbit every $\sim$130$P_\mathrm{b}$. Consequently, the oscillation reflects the behaviour of $r_{20\%}$ within one binary orbit. In the highly misaligned disc, the cavity is small enough to be influenced by the binary orbit, whereas no such oscillations appear in coplanar prograde or coplanar retrograde discs because their cavity sizes are larger. The coplanar retrograde disc has a smaller $r_{20\%}$ than the coplanar prograde disc but larger than polar discs, showing a gradual outward expansion. In the $i_0=150^\circ$ scenario, the disc experiences strong warping between 400 $P_\mathrm{b}$ and 1300 $P_\mathrm{b}$. During this warping phase, the low-density cavity expands quickly, becoming larger than that of any other configuration, as depicted in Figure~\ref{fig:surface}. However, since the inner disc fails to break and form a new inner polar disc, the cavity shrinks markedly after 1300 $P_\mathrm{b}$, becoming smaller than the coplanar prograde case. Like the coplanar retrograde disc, its $r_{20\%}$ also gradually expands outward over time.
	
	\subsubsection{Disc breaking}
	We focus on discs that undergo breaking. In our simulations, disc breaking occurs when $i_0\in\{45^\circ,\ 135^\circ\}$ for $q_\mathrm{b}\in\{0.67,\ 0.43\}$, and $i_0=150^\circ$ for $q_\mathrm{b}\in\{0.43,\ 0.25\}$. Given the morphological similarities observed in these discs \citep{smallwood2025}, we use the case of $i_0=45^\circ$ for $q_\mathrm{b}=0.67$ to illustrate the general features.

    \begin{figure*}
		\centering
		\begin{subfigure}[b]{0.6\textwidth}
			\centering
			\includegraphics[width=\textwidth]{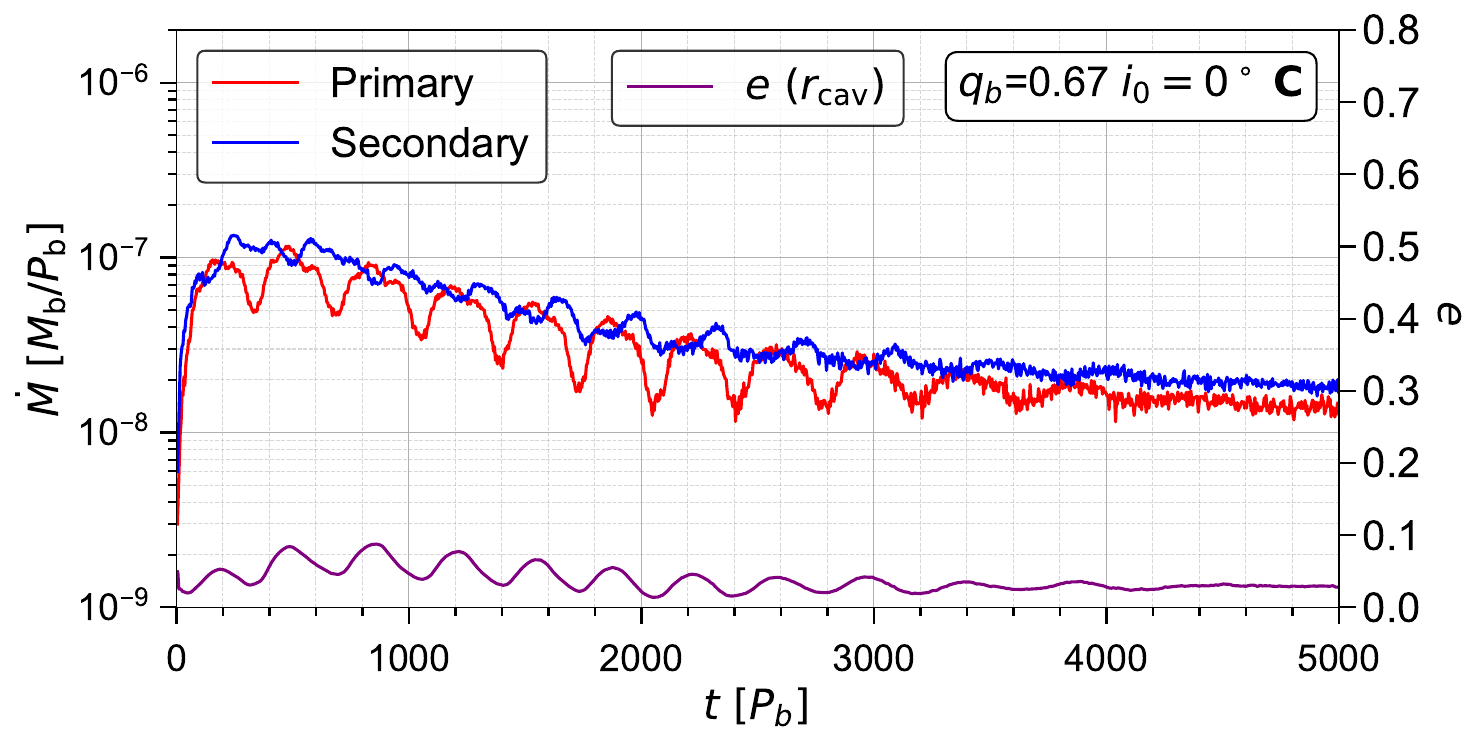}
			\label{fig:0p67i0_mdot}
		\end{subfigure}
		\\\vspace{-0.4cm}
		\begin{subfigure}[b]{\textwidth}
			\centering
			\includegraphics[width=\textwidth]{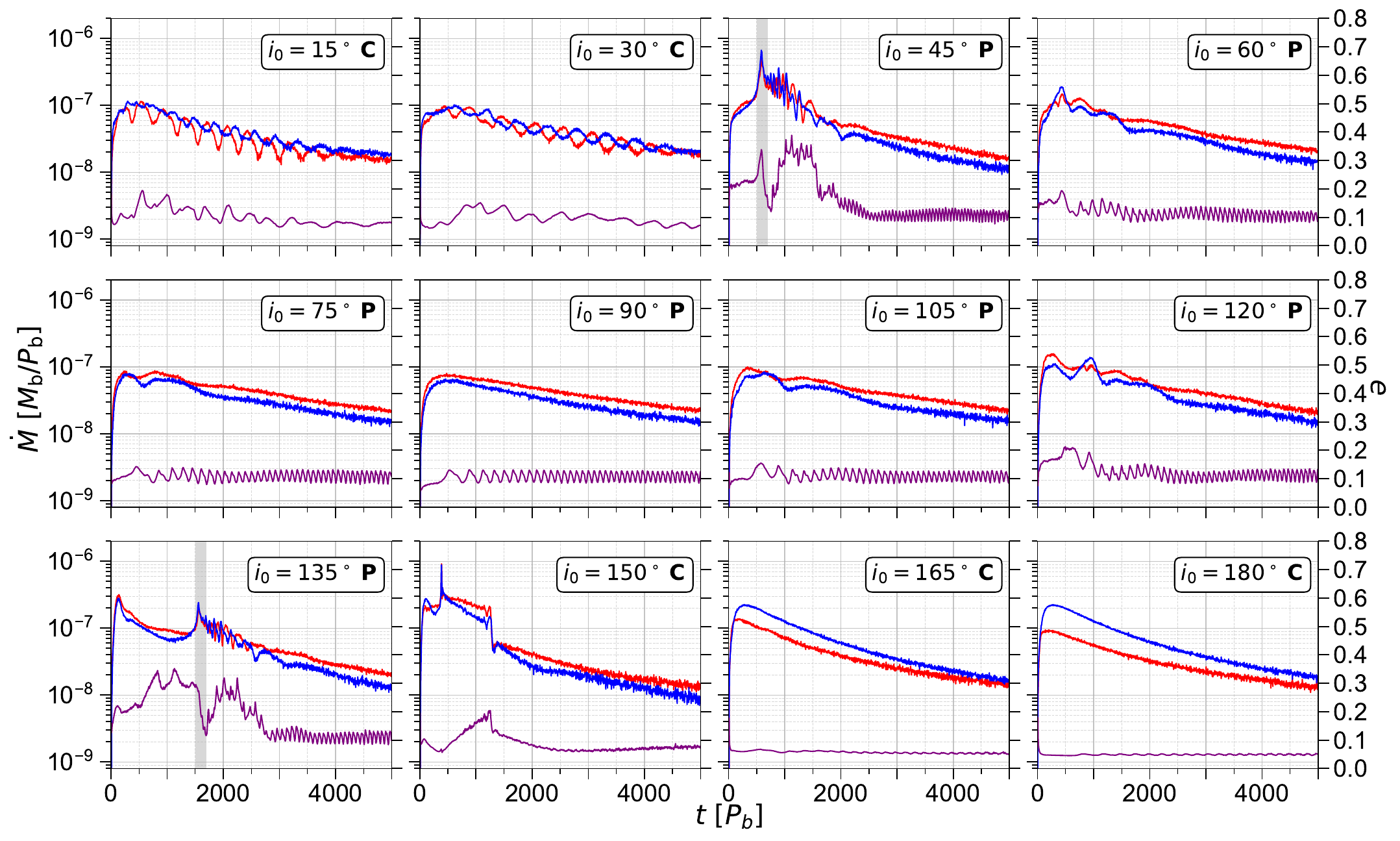}
			\label{fig:0p67_mdot}
		\end{subfigure}
		\caption{The accretion rate $\dot{M}_\mathrm{i}$ (where $\mathrm{i}=1,2$ correspond to primary and secondary), and cavity eccentricity $e\ (r=r_\mathrm{cav})$, as a function of time in units of initial binary orbital period $P_\mathrm{b}$, for $q_\mathrm{b}=0.67$ and different initial binary-disc tilts. Red and blue curves use the left $y$-axis, representing the primary and secondary accretion rates, respectively. Purple curves use the right $y$-axis, denoting the cavity eccentricity. The letters ``C'' or ``P'' correspond to whether the disc is undergoing coplanar or polar alignment, respectively.}
		\label{fig:mdot_0p67}
	\end{figure*}

    \begin{figure}
		\centering
		\includegraphics[width=\columnwidth]{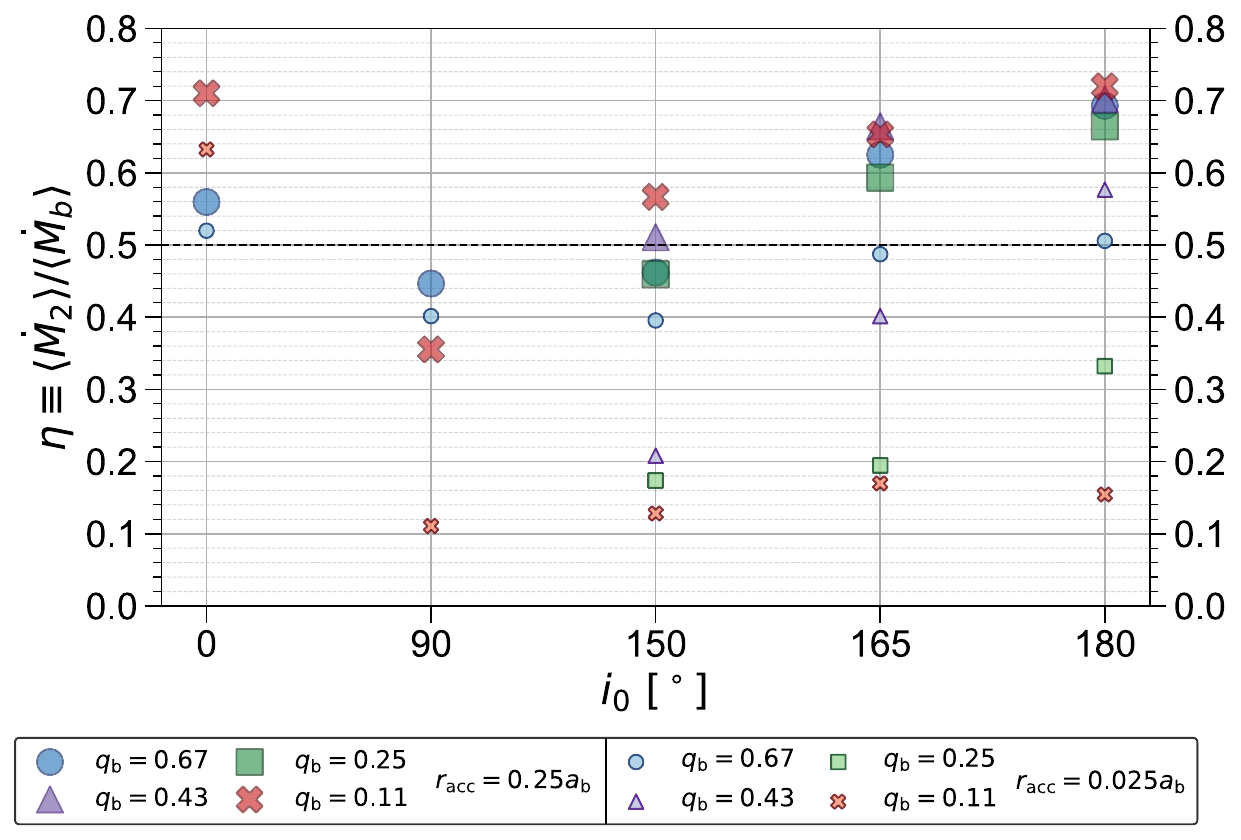}
		\caption{Comparison of the preferential accretion ratio $\eta = \langle \dot{M}_2 \rangle / \langle \dot{M}_\mathrm{b} \rangle$ for two accretion radii. The larger and smaller scatters correspond to simulations with $r_\mathrm{acc}=0.25$ and $r_\mathrm{acc}=0.025$, respectively. Note that for each initial condition, values of $\eta$ for both accretion radii are averaged over the simulation duration of the $r_\mathrm{acc}=0.025a_\mathrm{b}$ runs.}
		\label{fig:compare_eta}
	\end{figure}
	
	\begin{figure*}
		\centering
		\includegraphics[width=1.5\columnwidth]{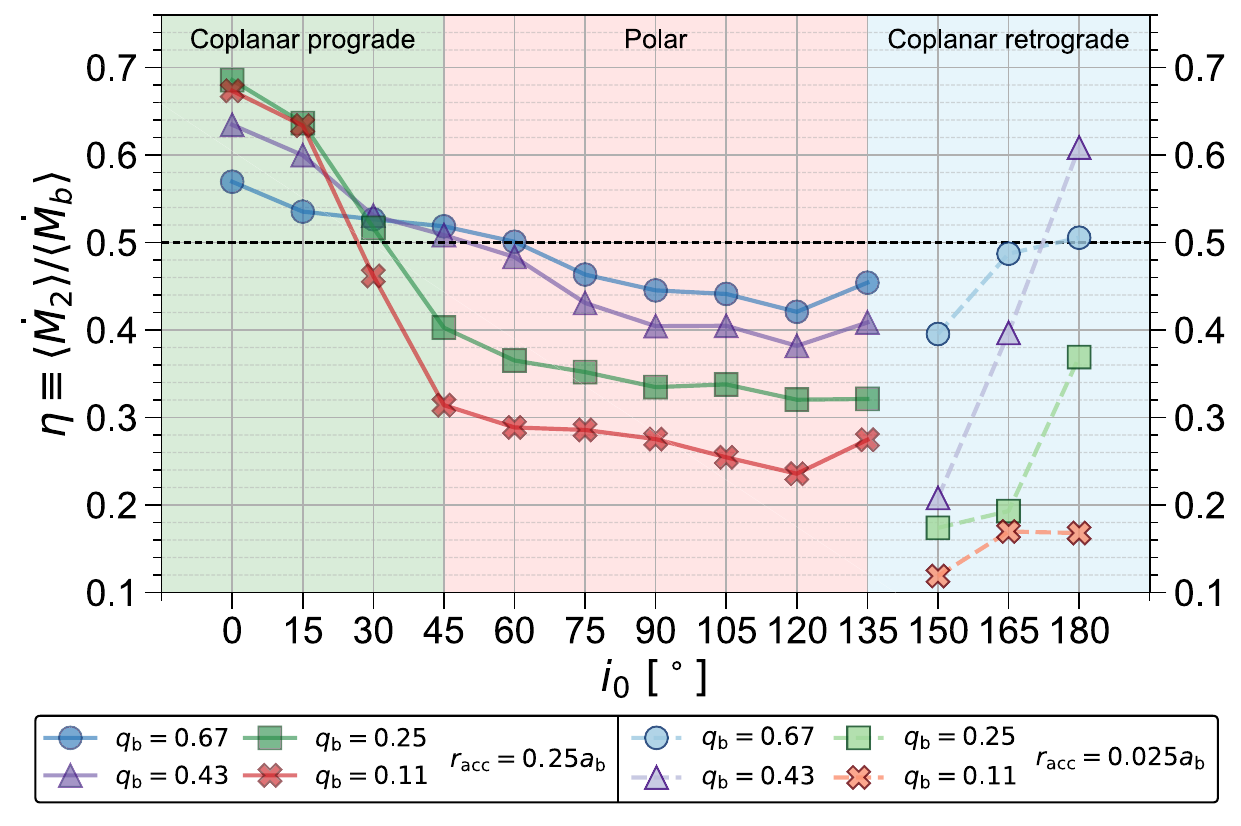}
		\caption{Preferential accretion ratio $\eta = \langle \dot{M}_2 \rangle / \langle \dot{M}_\mathrm{b} \rangle$ as a function of initial binary-disc tilt $i_0$ for four different binary mass ratios, denoted by different colours. Background shading marks the types of alignment. Solid and dotted lines correspond to simulations with accretion radii of $r_\mathrm{acc}=0.25a_\mathrm{b}$ and $0.025a_\mathrm{b}$, respectively. For the $r_\mathrm{acc}=0.25a_\mathrm{b}$ runs, the averaging time for $\eta$ is chosen separately for each mass ratio, according to their different alignment timescales, so the discs are compared at the similar stages of their tilt oscillations. The same averaging time is used across all $i_0$ for a given mass ratio. For the $r_\mathrm{acc}=0.025a_\mathrm{b}$ runs, the averaging time for each mass ratio is set by the duration of the shortest run due to the limited run time, and is likewise applied uniformly to all $i_0$ within that mass ratio.}
		\label{fig:eta}
	\end{figure*}
    
	Figure~\ref{fig:dens_evo_0p67i45} shows the evolution of the surface density from $t=500P_\mathrm{b}$ to $t=2000P_\mathrm{b}$. Before disc breaking, the whole disc is still precessing about the binary eccentricity vector, with the inner region exhibiting significant warping. During $500-800P_\mathrm{b}$, a newly formed inner disc tears, with angular momentum vector aligning with the binary's eccentricity vector. In contrast, the outer disc persists in precessing and gradually changes its longitude of the ascending node. The inner polar disc, initially eccentric and showing spiral density waves, gradually spreads outwards. The innermost region of this polar disc undergoes circularisation, eventually settling into a nearly circular configuration by $t=2000P_\mathrm{b}$. At this stage, the whole disc stabilises into polar alignment.
	
	Figure~\ref{fig:break_evo} depicts the evolution of the key disc properties during the disc breaking. Different coloured curves, ranging from dark red to ocean blue, represent distinct time snapshots from $t=500P_\mathrm{b}$ to $1200P_\mathrm{b}$, in intervals of $100P_\mathrm{b}$. The warp amplitude $\psi$ and surface density $\Sigma$ reveal that the break gradually propagates outward, creating a lower-density region in its vicinity. Within the inner disc, defined by radii smaller than the break radius, the eccentricity grows from approximately $0.25$ to $0.35$. Additionally, as the tilt of inner disc approaches $90^\circ$, its longitude of the ascending node evolves toward $90^\circ$, indicating that the inner polar disc is precessing about the binary eccentricity vector.
	
    \subsection{Long-term accretion variability}
    \label{subsec:accretion}
    
    We examine the long-term accretion features of unequal-mass binary systems with tilted circumbinary discs. The analysis of the disc morphology we performed in the previous section allows us to establish a correlation between the global disc structure and the temporal evolution of the binary accretion rates, providing insights into how the disc morphology influences the mass transfer processes within these systems. 
    
    Figure~\ref{fig:mdot_0p67} shows the long-term accretion rate, $\dot{M}$, and the cavity eccentricity, $e$, as a function of time for $q_\mathrm{b}=0.67$. We first focus on the case of a coplanar prograde disc ($i_0 = 0^\circ$), shown in the first panel. The red and blue curves represent the accretion rates of the primary and secondary stars, respectively, whose values are shown by the left $y$-axis. The purple curve shows the eccentricity of the cavity, whose values are shown by the right $y$-axis. We can see that the primary and secondary accrete alternately, so there is a temporary preferential accretion on one of the two components. This can be attributed to the disc apsidal precession, where the inner disc region becomes eccentric and slowly precesses. For a coplanar pressure-less particle disc, its secular apsidal precession rate around an eccentric binary is given by \citet{munoz2016a}
	\begin{equation}
		\dot{\omega}_\mathrm{d}\approx\frac{3\Omega_\mathrm{b}}{4}\frac{q_\mathrm{b}}{(1+q_\mathrm{b})^2}\left(1+\frac32 e^2_\mathrm{b}\right)\left(\frac{a}{r}\right)^{7/2},
		\label{eq:apsidal_precession}
	\end{equation}
    the apsidal precession timescale at $r=3.5a_\mathrm{b}$ is $\sim324P_\mathrm{b}$. In an unequal-mass binary system, the secondary star occupies a wider Keplerian orbit than its primary star. This wider orbit allows the secondary star to accrete more material around apastron, resulting in a higher average accretion rate compared to the primary star \citep{munoz2020a}.
    
    However, when the disc is misaligned, the primary star is closer to the inner disc during an orbital period and has a larger accretion sphere. The secondary star's ability to capture gas around apastron compared to the coplanar scenario is also hindered. This effect becomes more pronounced with increasing misalignment. Consequently, as $i_0$ increases from $0^\circ$ to $30^\circ$, the accretion rates of the primary and secondary become more comparable, resembling those of an equal-mass binary in a coplanar disc.
	
    In the cases of discs evolving toward polar alignment ($38.5^\circ\le i_0\le 141.5$), we can see the primary star dominates the accretion as $i_0$ is closer to $90^\circ$. Similar to previous studies \citep[e.g.,][]{nixon2013,smallwood2025}, the closer $i_0$ is to $i_\mathrm{lower}$ and $i_\mathrm{upper}$, the more eccentric and warped the inner disc becomes. For nearly polar discs, no preferential alternating accretion is seen, because the binary gravitational torque cannot efficiently excite the apsidal precession of the inner disc. The oscillation timescale of the cavity eccentricity matches that of the cavity size (see Section~\ref{subsubsec:cavity}). For $i_0=45^\circ$ and $i_0=135^\circ$, as the disc is strongly warped and the eccentricity of the cavity pumps, the binary accretes quickly before breaking. During the breaking, the inner polar disc forms (see Figure~\ref{fig:dens_evo_0p67i45}) and the cavity eccentricity initially decreasing to approximately 0.15 before rising to 0.35. The newly formed disc precesses about the binary eccentricity vector, and undergoes apsidal precession, leading to the preferential alternating accretion. As the break propagates outward, the inner disc's eccentricity gradually dampens, reducing the magnitude of the alternating accretion over time. Eventually, the inner disc becomes circularised, causing the alternating accretion phenomenon to cease. As a result, the entire disc settles into a polar alignment, where its angular momentum vector aligns with the eccentricity vector of the binary system. In this final state, the accretion behaviour closely resembles that of a disc that was initially polar aligned.

    In cases where the disc evolves into a coplanar retrograde alignment ($i_0 \geq 150^\circ$), there is no preferential alternating accretion and no pronounced oscillation of the cavity eccentricity. This is because the angular momentum vector of the binary is opposite to that of the disc, preventing the presence of prograde resonances \citep[e.g.,][]{li2021a}. Retrograde discs generally accrete a greater amount of mass compared to their prograde counterparts \citep{tiede2023}. In prograde systems, resonances act to keep the gas away from the binary region, reducing the overall accretion rate. Similar to the coplanar prograde cases, the secondary component of the binary in a retrograde orbit still accretes more mass than the primary if the initial tilt of the disc is higher than $150^\circ$.
	
    We now focus on the effect of the binary mass ratio. We list the long-term accretion results for the other three mass ratios in Appendix~\ref{appendix:Accretion} (Figure~\ref{fig:mdot_0p43}, \ref{fig:mdot_0p25}, \ref{fig:mdot_0p11}). With smaller mass ratio, the apsidal precession timescales (equation~\ref{eq:apsidal_precession}) in nearly coplanar prograde discs and the accretion oscillation timescales in warped discs both increase. Furthermore, for the small mass ratios $q_\mathrm{b}=0.25,\ 0.11$, the primary star in nearly coplanar retrograde discs ($i_0=150^\circ,\ 165^\circ$) finally dominates the accretion. Only in nearly coplanar prograde discs ($i_0\le15$) or in coplanar retrograde discs ($i_0=180^\circ$) does the secondary dominate accretion.

    In prior simulations, we employed a binary accretion radius of 0.25$a_\mathrm{b}$ to optimise the computational efficiency. However, this choice may result in overestimated accretion rates, especially in cases with smaller binary mass ratio and retrograde discs \citep{amaro-seoane2016}. To investigate the impact of the accretion radius on preferential accretion, we performed simulations with accretion radius $r_\mathrm{acc}=0.025 a_\mathrm{b}$, initial binary-disc tilts $i_0\in\{0^\circ,\ 90^\circ,\ 180^\circ\}$ and mass ratios $q_\mathrm{b}\in\{0.67,\ 0.11\}$. Due to the significantly increased computational cost, coplanar discs were simulated up to 2000$P_\mathrm{b}$, whereas polar discs were simulated up to 1000$P_\mathrm{b}$.
    
    Figure~\ref{fig:compare_eta} shows the comparison of preferential accretion ratio $\eta = \langle \dot{M}_2 \rangle / \langle \dot{M}_\mathrm{b} \rangle$ for two accretion radii. For each initial condition, $\eta$ for both accretion radii is averaged over the simulation duration of the corresponding $r_\mathrm{acc}=0.025a_\mathrm{b}$ run. Larger and smaller scatters represent the data from $r_\mathrm{acc}=0.25$ and $r_\mathrm{acc}=0.025$, respectively. Details of the binary accretion rates for the smaller accretion radius are shown in Figure~\ref{fig:compare_racc}. When the accretion radius is reduced from a larger to a smaller value, for $q_\mathrm{b}=0.67$, $\eta$ in coplanar prograde and polar discs decrease by no more than 8\%. In coplanar retrograde discs, $\eta$ decreases from 0.7 to 0.51 with decreasing accretion radius. For $q_\mathrm{b}=0.11$, $\eta$ in the coplanar prograde disc decreases by less than 10\%, while in the polar disc case it drops from 0.36 to 0.11 using a smaller accretion radius. The accretion patterns in both coplanar prograde and polar discs remain consistent across the two accretion radii. The only significant effect of the accretion radius is on the preferential accretion experienced by the binary in the coplanar retrograde disc where $\eta$ decreases from 0.72 to 0.15, reversing the preferential accretion from the secondary to the primary.

   In order to determine whether decreasing the accretion radius causes preferential accretion to switch from the secondary to the primary for discs evolving towards coplanar retrograde alignment, we performed additional simulations with $r_\mathrm{acc}=0.025a_\mathrm{b}$, initial tilts $i_0\in\{150^\circ,\ 165^\circ,\ 180^\circ\}$, and all mass ratios. Most runs were evolved to 2000$P_\mathrm{b}$, though some were run up to 1000$P_\mathrm{b}$ or 500$P_\mathrm{b}$ due to limited computational resources. The detailed setups are listed in Table~\ref{tab:setup}. Their $\eta$ values are shown in Figure~\ref{fig:compare_eta}. Except for the cases with $q_\mathrm{b}\in\{0.67,\ 0.43\}$ at $i_0=180^\circ$, where the secondary still dominates accretion regardless of the accretion radius, all other cases show primary dominance. This suggests that for discs evolving towards coplanar retrograde alignment at low binary mass ratio, the preferential accretion generally occurs on the primary rather than secondary. We further discuss this finding in Section~\ref{subsec: accretion radius}.
    
    Since the preferential accretion behaviour remains robust across two accretion radii in discs evolving toward coplanar prograde and polar alignments, we are able to investigate how the initial tilt affects $\eta$ for each mass ratio, and how the binary mass ratio affects $\eta$ for each initial tilt. We account for the mass ratio dependence of the alignment timescale using equation~\ref{eq:Omega_d} and adopt a fixed fraction of the nodal precession timescale as averaging times to average $\eta$ for all mass ratios. This corresponds to $692\,P_\mathrm{b}$, $904\,P_\mathrm{b}$, $1560\,P_\mathrm{b}$ and $5000\,P_\mathrm{b}$ for $q_\mathrm{b}=0.67,\,0.43,\,0.25,\,0.11$, respectively. This ensures that, for a given initial tilt, discs with different mass ratios follow nearly identical tilt oscillation, allowing us to isolate the effect of the mass ratio. Meanwhile, by using a fixed time for averaging $\eta$ across all discs at a given mass ratio, we can examine the effect of the initial tilt. The resulting $\eta$ values are shown in Figure~\ref{fig:eta}.
    
    For each mass ratio, $\eta$ generally decreases as $i_0$ increases from $0^\circ$ to $120^\circ$. This trend suggests that, as the initial tilt increases, accretion becomes increasingly dominated by the primary. Focusing instead on the mass ratio dependence, we find that for discs with $i_0\le15^\circ$, $\eta$ increases as $q_\mathrm{b}$ decreases from 0.67 to 0.25, whereas for discs with $i_0\ge30^\circ$, $\eta$ generally decreases as $q_\mathrm{b}$ decreases from 0.67 to 0.11. Figure~\ref{fig:eta} also presents results from simulations with small accretion radius for discs evolving towards coplanar retrograde alignment. For each mass ratio, $\eta$ is averaged over the same time interval, using the duration of the shortest simulation. Preferential accretion onto the secondary appears only in the cases with $q_\mathrm{b}\in\{0.67,\ 0.43\}$ at $i_0=180^\circ$. For mass ratios $q_\mathrm{b}=0.67$, 0.43 and 0.25, $\eta$ shows an increasing trend with increasing initial tilt.
    
    \subsection{Disc mass evolution}
	
    \begin{figure}
        \centering
        \includegraphics[width=\linewidth]{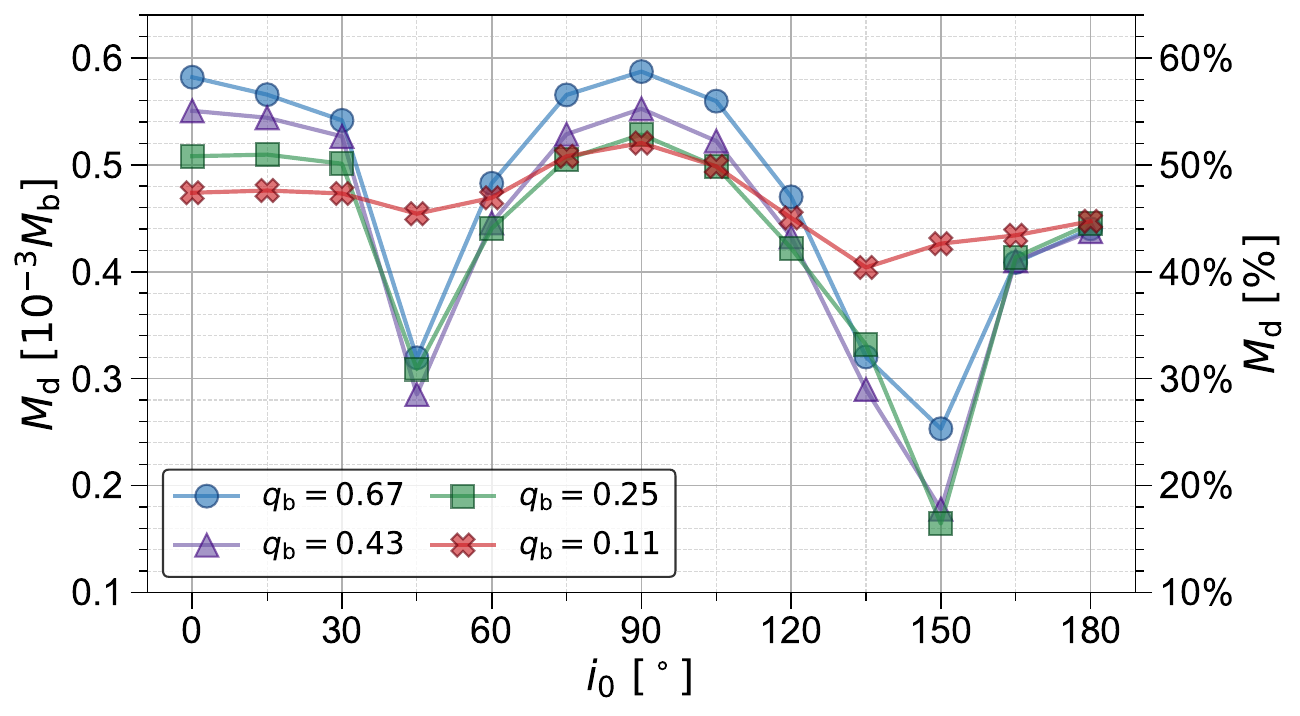}
        \caption{Disc mass $M_\mathrm{d}$ as a function of initial binary-disc tilt $i_0$ at 5000 binary orbital periods for four distinct mass ratios, represented by varying colours. The left $y$-axis is normalised to the total disc mass, while the right $y$-axis is expressed as a percentage.}
        \label{fig:disc_mass}
    \end{figure}
    
    Figure~\ref{fig:disc_mass} shows the disc mass as a function of initial binary-disc tilt $i_0$ at 5000$P_\mathrm{b}$ for four mass ratios. For $q_\mathrm{b}\ge0.25$, discs with $i_0=150^\circ$ experience a significant reduction of mass, retaining only about $17$ per cent of their initial mass for $q_\mathrm{b}=0.43,\, 0.25$, and $25$ per cent for $q_\mathrm{b}=0.67$. Discs with $i_0 = 45^\circ$ and $135^\circ$ retain about $30$ per cent of their initial mass. These discs' initial tilts lie close to the tilt criterion, where the discs undergo strong warping or even breaking, leading to the quick depletion of the disc mass, which can potentially impede the process of planet formation.
    
    For $q_\mathrm{b}=0.11$, the mass loss rates of discs with $i_0$ around $45^\circ$ and $135^\circ$ are lower than those of discs with other mass ratios. Within the simulated time, these discs do not experience strong warping or breaking and therefore retain more disc mass. Nevertheless, their mass loss rates remain higher than those of discs at other initial tilts within the same mass ratio. For the remaining cases, discs retain 40 per cent to 60 per cent of their initial mass. Notably, the mass loss rates of discs with $i_0 = 90^\circ$ are the lowest across all mass ratios, slightly lower than those of discs with $i_0 = 0^\circ$. Retrograde discs exhibit higher mass loss compared to their prograde counterparts, which is consistent with \citet{nixon2013}, indicating that weaker resonances in retrograde discs enable accretion of gas particles more easily.
    
    It is noteworthy that the numerical dissipation might contribute to an overestimation of the mass loss rate. Despite these limitations, the general trend is expected to be largely robust. The relationship between disc mass loss rate and initial tilt, provides valuable insights into the dynamics of circumbinary disc evolution and the potential challenges for circumbinary planet formation in such environments.

	\section{Discussion}
	\label{sec: discussion}
    \subsection{Effect of the accretion radius}
    \label{subsec: accretion radius}
    In the {\sc phantom} code, a gas particle is accreted without any additional checks if it falls within $0.8r_\mathrm{acc}$. In the region $0.8r_\mathrm{acc}\le r \le r_\mathrm{acc}$, a gas particle undergoes accretion by a sink particle if it passes accretion checks \citep{price2018}. Our choice of $r_\mathrm{acc}=0.25a_\mathrm{b}$ may not be optimal for small mass ratios and retrograde discs, as it exceeds the bound radius of each sink particle.

    Figure~\ref{fig:compare_eta} shows the preferential accretion ratio $\eta$ from simulations with $r_\mathrm{acc}=0.25a_\mathrm{b}$ and $r_\mathrm{acc}=0.025a_\mathrm{b}$, while Figure~\ref{fig:compare_racc} presents the corresponding binary accretion rates for $r_\mathrm{acc}=0.025a_\mathrm{b}$. For cases with $i_0\ge150^\circ$, adopting a smaller accretion radius leads to a significant drop of $\eta$, reversing the preferential accretion from secondary onto primary, except for initially coplanar retrograde discs $q_\mathrm{b}\in{0.67,\,0.43}$.
    
    Our results using the larger accretion radius are qualitatively similar to those of \citet{nixon2015}, who adopted $r_\mathrm{acc}=0.2a_\mathrm{b}$ and found that the secondary accreted nearly all the gas in coplanar retrograde discs around eccentric binaries with low mass ratios. However, these results likely overestimate secondary accretion in coplanar retrograde discs. 
    Indeed, the accretion radius plays an important role in retrograde discs as counter-rotation of gas particles relative to the binary results in higher relative velocities, thereby reducing the interaction region \citep{amaro-seoane2016}. Furthermore, the absence of gas streams between the binary components in retrograde discs likely suppresses any exchange of mass \citep{dittmann2025a}. The total binary accretion rates remain instead nearly identical between our large and small accretion radius simulations, suggesting that the discrepancies in $\eta$ primarily arise because gas particles more bound to the primary or not bound to the secondary are erroneously accreted by the secondary, and vice versa.
    
    Our simulations with a small accretion radius indicate that, for low mass ratios, the primary will generally dominate the accretion in discs evolving towards coplanar retrograde alignment. This agrees with previous studies reporting the preferential accretion onto the primary in coplanar retrograde discs \citep{bourne2024,dittmann2025a}. The exceptions for initially coplanar retrograde discs with $q_\mathrm{b} \in {0.67, 0.43}$ suggest that there may exist a mass ratio range between 1 and 0.25 where the secondary can dominate accretion in coplanar retrograde discs. However, a detailed investigations of this regime, including the precise measurements of $\eta$ and the effect of initial conditions, would require a broader parameter space and is beyond the scope of this study.
    
	\subsection{Implications}
	\subsubsection{Observations}
	\label{subsec: observation}

    For the detection of binary accretion processes, high-resolution spectrographs such as VLT SPHERE/ZIMPOL \citep{schmid2018}, HST COS \citep{green2011}, and VLT X-shooter \citep{vernet2011,rigliaco2012} are invaluable. These instruments cover the UV, optical, and Near-IR wavelengths, enabling the measurement of accretion diagnostics like H$\alpha$, Ca II, and He I emission lines. These diagnostics are crucial for determining individual star accretion rates within binary systems. Additionally, X-ray observatories such as \textit{Chandra}/HETGS \citep{kastner2002} and XMM-Newton/RGS \citep{stelzer2004} offer further insights into magnetospheric accretion and shocks, allowing for the distinction of hotspots on individual components of a binary system \citep{gunther2007}.
        
    On the other hand, resolving the detailed structure of circumbinary discs necessitates facilities with high angular resolution and sensitivity at longer wavelengths. The Atacama Large Millimeter/submillimeter Array (ALMA) \citep{almapartnership2015} has been revolutionary in this regard, providing detailed images of circumbinary discs. These images reveal features such as cavities and spiral arms, which trace the dynamics of gas and dust within these systems. Furthermore, the Very Large Telescope Interferometer (VLTI), equipped with instruments like GRAVITY \citep{abuter2017} and MATISSE \citep{lopez2022}, has facilitated near- and mid-infrared interferometric observations. These observations provide insights into the inner regions of circumbinary discs and the interactions between the disc and the binary stars. A comprehensive summary of information on circumbinary systems can be found in \citet{czekala2019}.

    A notable example of a binary system exhibiting preferential accretion within a polar circumbinary disc is HD 98800 B. HD 98800 is a quadruple system located at a distance of 47 pc, consisting of two binaries, AaAb and BaBb \citep{leeuwen2007}. The orbit of the binary BaBb is well constrained with a semi-major axis $a_\mathrm{BaBb} = 1$ AU and an eccentricity $e_\mathrm{BaBb} = 0.7849 \pm 0.0053$. The mass ratio of BaBb is 0.833, with components masses of $M_\mathrm{Ba} = 0.699 \pm 0.064M_\odot$ and $ M_\mathrm{Bb} = 0.582 \pm 0.051 M_\odot$ \citep{boden2005}. Based on ALMA observations, \citet{kennedy2019a} inferred the binary-disc tilt of the binary HD 98800 B is either $48^\circ$ or $90^\circ$, with the latter being more likely as the polar alignment timescale is shorter than the system's age. Using the SPHERE instrument \citep{beuzit2019} at the Very Large Telescope (VLT), \citet{zurlo2020} observed HD 98800 BaBb with the H$\alpha$ filter of ZIMPOL \citep{schmid2018}. They detected H$\alpha$ flux from the primary star but none from the secondary. This observation is consistent with our simulations, which show that a binary within a circumbinary disc evolving into a polar configuration exhibits preferential accretion onto the primary star. This result also aligns with the SPH numerical findings of \citet{smallwood2022}.

	\subsubsection{Formation of circumbinary planets}

    Circumbinary discs that experience breaking and retrograde motion tend to lose mass at a more rapid pace compared to those that maintain coplanar or nearly polar alignments. The dynamical interactions within these discs can significantly influence the conditions necessary for planet formation. Specifically, if a planet forms within a tilted, massive retrograde circumbinary disc, the swift depletion of disc mass may result in the planet becoming decoupled from the disc, as the Stokes number of a particle is inversely proportional to the gas disc surface density. This decoupling can lead to the planet adopting a misaligned orbit relative to the binary system, thus becoming a misaligned circumbinary planet \citep{martin2019}.

    Conversely, coplanar and polar discs are more conducive to providing stable environments for planet formation. These discs are better able to retain a significant portion of their mass over the simulated timescales, which is crucial for sustained planetesimal growth and the eventual formation of planets. 99 Herculis is a binary star system with an estimated age $6\sim10$ Gyr \citep{nordstrom2004,takeda2007} and hosts a circumbinary debris disc \citep{kennedy2012,smallwood2020}. The remarkable longevity of this debris disc may be attributed to the relatively low mass loss rate in polar discs. Furthermore, differential nodal precession between gas and dust in circumbinary discs evolving towards polar alignment can create dust traffic jams, leading to enhanced dust concentrations that favour the formation of polar circumbinary planets \citep{smallwood2024a,smallwood2024b}. However, due to current selection effects and observational limitations, the number of confirmed polar circumbinary planets remains zero. This scarcity is partly due to the challenges associated with detecting planets in such configurations, as they may not exhibit the same observable signatures as planets in a more common, coplanar alignment. There is, however, indirect evidence of a polar planet around the post–asymptotic giant branch (AGB) star binary AC Her \citep{martin2023b}. More recently, \citet{baycroft2025a} presented evidence for a polar circumbinary exoplanet in system 2M1510, which hosts a binary consisting of two equal-mass brown dwarfs with an eccentricity of 0.36.
    
	\subsection{Disc breaking}

	Generally, there are two regimes describing warped disc, depending on the relation between disc aspect ratio $H/r$ and its viscosity coefficient $\alpha_\mathrm{SS}$ \citep{shakura1973}. When $\alpha_\mathrm{SS}>H/r$, the disc is in the diffusive regime, which usually occurs in fully-ionised and thin discs, such as those orbiting binary black holes. When $\alpha_\mathrm{SS}<H/r$, the disc is in the bending-wave regime, which often applies to the protoplanetary discs, where the warp propagates like a wave through the disc at half the sound speed $c_\mathrm{s}/2$ \citep{papaloizou1995,lubow2000}. In our work, the discs are in the bending-wave regime, as $\alpha_\mathrm{SS}=0.01<H/r=0.1$. For a tilted circumbinary disc, the inner regions nodally precess faster than the outer regions. Therefore, a tilted disc may develop a warp when the outer regions of the disc are radially extended significantly. When the disc undergoes strong warping, disc breaking can occur, splitting the disc into separate rings.
    
        \begin{figure}
            \centering
            \includegraphics[width=\linewidth]{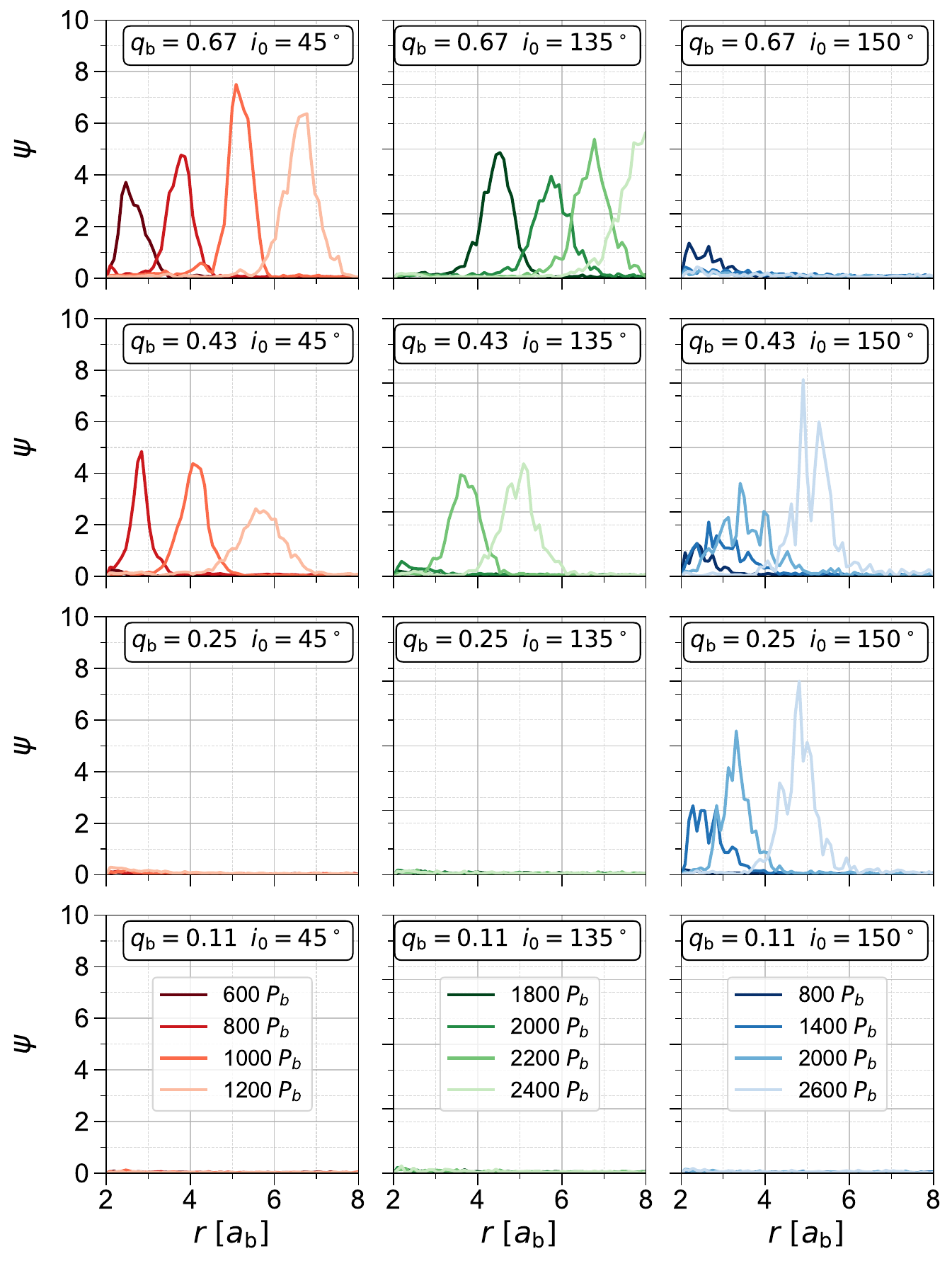}
            \caption{The evolution of disc warp amplitude $\psi$ as a function of radius $r$ for initial binary-disc tilts $i_0 = 45^\circ$, $135^\circ$, and $150^\circ$ across four binary mass ratios, $q_\mathrm{b} = 0.67, 0.43, 0.25, 0.11$. Note that we select different time intervals for each tilt.}
            \label{fig:warp_amplitude}
        \end{figure}
        
	Figure~\ref{fig:warp_amplitude} illustrates the evolution of warp amplitude. The different rows represent different mass ratios: $q_\mathrm{b}=0.67$, $0.43$, $0.25$ and $0.11$ (top to bottom). The different columns refer to different initial binary-disc tilts: $i_0=45^\circ$, $135^\circ$ and $150^\circ$ (left to right). First, we study the effect of the binary mass ratio. The most significant change due to the mass ratio is the disc precession timescale (e.g. equation~\ref{eq:Omega_d}, ~\ref{eq:k}). We can see the break time is later when mass ratio gets smaller. For the $i_0=45^\circ$ or $i_0=135^\circ$ cases, as the mass ratio becomes smaller, the warp amplitude decreases. This phenomenon aligns with previous analytical and numerical studies \citep{facchini2013,young2023}. However, for the $i_0=150^\circ$ cases, the disc does not break for $q_\mathrm{b}=0.67$, showing only moderate warping. The warp amplitudes for $q_\mathrm{b}=0.43$ and $0.25$ are quite similar, suggesting that the warp amplitude or break criteria are also affected by other properties.
    
    Second, we investigate the impact of misalignment between the binary and the circumbinary disc. In our study, all broken discs are found to be in retrograde orientations. At the point of disc breaking, the binary-disc tilt ranges from $105^\circ$ to $150^\circ$. We do not observe any disc breaking when the tilt is around $45^\circ$. This difference may be attributed to the fact that in retrograde discs, the inner regions of the disc are closer to the binary system \citep{nixon2011,nixon2015}. As a result, these inner regions experience a stronger gravitational torque from the binary compared to the prograde configuration.
	
    We acknowledge that numerous additional parameters can influence the criteria for disc breaking. \citet{rabago2024} discovered that discs with smaller inner cavities, cooler temperatures, and steeper power-law profiles are more susceptible to breaking. \citet{deng2022} demonstrated that broken discs can potentially reconnect in adiabatic simulations. Predicting the break radius and time is a critical area of study, as it can provide valuable insights for interpreting and forecasting observations of disc breaking. In the bending wave regime, \cite{deng2022} showed that arbitrary warps around a central potential with a quadruple moment tend to evolve toward uniformly precessing states. Such states cease to exist when the outer disc misalignment exceeds $\sim 20$ degrees, leading to disc breaking, as verified by direct numerical simulations. However, the potential around an eccentric binary is more complicated, and the disc considered here is thicker than \citet{deng2022}. Discs in the diffusive regime exhibit distinct behaviour, as warps propagate through them in a diffusive manner. However, previous analytical studies of break criteria in the diffusive regime \citep{dogan2018, dogan2020} were derived for steady warps and may not be directly applicable to simulations, where the disc structure is continuously evolving \citep{young2023,rabago2024}. These considerations emphasise the need for a more comprehensive understanding of the various factors that contribute to disc breaking, as well as the development of updated analytical models that can accurately capture the dynamic nature of evolving disc structures.

	\subsection{Numerical limitations}
	\label{subsec: Numerical limitations}

    Our study described above incorporates certain simplifications. Primarily, our simulations adopt a locally isothermal equation of state. This approach assumes that heat generated by viscosity or shocks is rapidly dissipated through a cooling process, quickly restoring the disc temperature to equilibrium. However, this treatment may not accurately represent the behaviour of circumstellar discs, potentially impacting the binary orbital separation \citep{li2022} and individual accretion rates. Future studies should incorporate more physically realistic processes, such as heating and cooling mechanisms. These processes have been explored in 2D grid-based simulations \citep{wang2023b,wang2023}, providing valuable insights that can be integrated into our future models.

    Another significant limitation of this study is the insufficient resolution of the central binary region, which results from the extended simulation duration. While the circumbinary discs remain well-resolved throughout all simulations (see Appendix~\ref{appendix:Resolution}), the accretion flows and circumstellar discs around individual stars remain unresolved. This issue is a common challenge in SPH simulations, where resolving low-density regions is particularly difficult. Due to the limited resolution, the viscous timescale in these regions is artificially shortened, leading to a rapid accretion of material from the circumstellar discs onto individual binary stars. This may fail to account for the buffering effect of circumstellar discs on stellar accretion. For a nearly coplanar prograde disc around an eccentric binary, some gas particles may be ejected from the circumstellar discs and later return to the inner disc region. These particles can then interact with the inner region of the circumbinary disc \citep{tiede2022}. Future work can leverage the recently implemented adaptive particle refinement (APR) feature in the {\sc phantom} code \citep{nealon2024} to enhance the resolution within the binary cavity.

	\section{Summary}
	\label{sec: summary}
	
	We conducted a suite of 3D SPH simulations to investigate the dynamics of low-mass misaligned circumbinary discs around unequal-mass eccentric binaries. Our parameter study explored four binary mass ratios ($q_\mathrm{b}=0.67, 0.43, 0.25, 0.11$) with initial binary-disc tilts ($i_0$) ranging from $0^\circ$ to $180^\circ$. Our results show that both the mass ratio and the initial tilt alter disc evolution and long-term binary accretion variability. 
    
    In our simulations, discs evolving towards polar and coplanar retrograde generally favour accretion onto the primary star, while discs evolving towards coplanar prograde generally favour accretion onto the secondary. For discs with the same initial tilt, when compared at similar stages of their tilt oscillations, we find the preferential accretion ratio $\eta=\langle\dot{M_2}\rangle/\langle\dot{M_\mathrm{b}}\rangle$ shows a non-monotonic function of the mass ratio. $\eta$ increases as $q_\mathrm{b}$ decreases from 0.67 to 0.25 when $i_0 \le 15^\circ$. In contrast, for discs with $30^\circ \le i_0 \le 135^\circ$, $\eta$ decreases as $q_\mathrm{b}$ decreases from 0.67 to 0.11. Across all mass ratios, polar discs show the lowest mass loss rates, slightly below those of coplanar prograde discs, whereas retrograde discs tend to lose mass more rapidly than their prograde counterparts. Discs that undergo strong warping or breaking in our simulations experience rapid mass loss.

    This work provides insights into interpreting observations of accreting binaries in young stellar systems. The predicted connection between inner disc eccentricity and alternating accretion patterns may help explain observed variability in pre-main sequence binaries such as TWA 3A \citep{tofflemire2017b,tofflemire2019,anthonioz2015,czekala2021}. Our finding that the primary dominates the accretion in polar discs aligns with H$\alpha$ observations of HD 98800 B \citep{kennedy2019a}. These results offer a potential diagnostic for identifying binary-disc misalignment and hint that the observed demographics of circumbinary planets may reflect the history of their formation environment. However, further exploration of the parameter space, incorporating radiative transfer and higher-resolution simulations, will be needed to test the universality of these findings.

	\section*{Acknowledgements}
    We thank the anonymous referee for the insightful comments and suggestions. RY acknowledges the support and hospitality provided by the 2024 Protoplanetary Disc and Planet Formation Summer School and Workshop in Beijing. RY and S.-F.L. are supported by the Guangdong Basic and Applied Basic Research Foundation (grant No. 2021B1515020090) and the National Natural Science Foundation of China (grant No. 11903089). JLS acknowledges funding and support from the Dodge Family Prize Fellowship in Astrophysics and by the Vice President of Research and Partnerships of the University of Oklahoma and the Data Institute for Societal Challenges. Y-PL is supported in part by the Natural Science Foundation of China (grants 12373070, and 12192223), the Natural Science Foundation of Shanghai (grant NO. 23ZR1473700). We thank Daniel Price for providing the {\sc phantom} \citep{price2018} code for our SPH simulations and the {\sc splash} \citep{price2007} code for rendering some of our figures. This research was enabled in part by support provided by the High-performance Computing Platform of Peking University. Numerical simulations were conducted on the Loong cluster at the School of Physics and Astronomy, Sun Yat-sen University, China, and on the Béluga and Narval clusters at the Digital Research Alliance of Canada in Montreal, Canada.
	
	\section*{Data Availability}
	The input files used to generate the simulations will be shared on reasonable request to the corresponding author. The numerical simulations were performed using the public version of the {\sc phantom} \citep{price2018}. Visualisations of the discs were generated by the {\sc splash} \citep{price2007}. Figures were produced using {\sc matplotlib} \citep{hunter2007} and {\sc numpy} \citep{harris2020}.

	
	
	\bibliographystyle{mnras}
	\bibliography{cbd}
	
	
	
	
	\appendix
	
	\section{Resolution}
	\label{appendix:Resolution}
	
	In the SPH code {\sc phantom}, the viscosity is calculated using an artificial viscosity $\alpha_\mathrm{AV}$ (see equation~\ref{eq:viscosity}). As the simulation evolves, the average smoothing length per scale height, $\langle h\rangle/H$, typically increases and can vary significantly due to the warping and breaking of the disc. To accurately capture the wave-like evolution of the disc, high resolution is needed, typically requiring that $\langle h\rangle/H$ remains below 1.
    
\begin{figure}
		\centering
		\includegraphics[width=\linewidth]{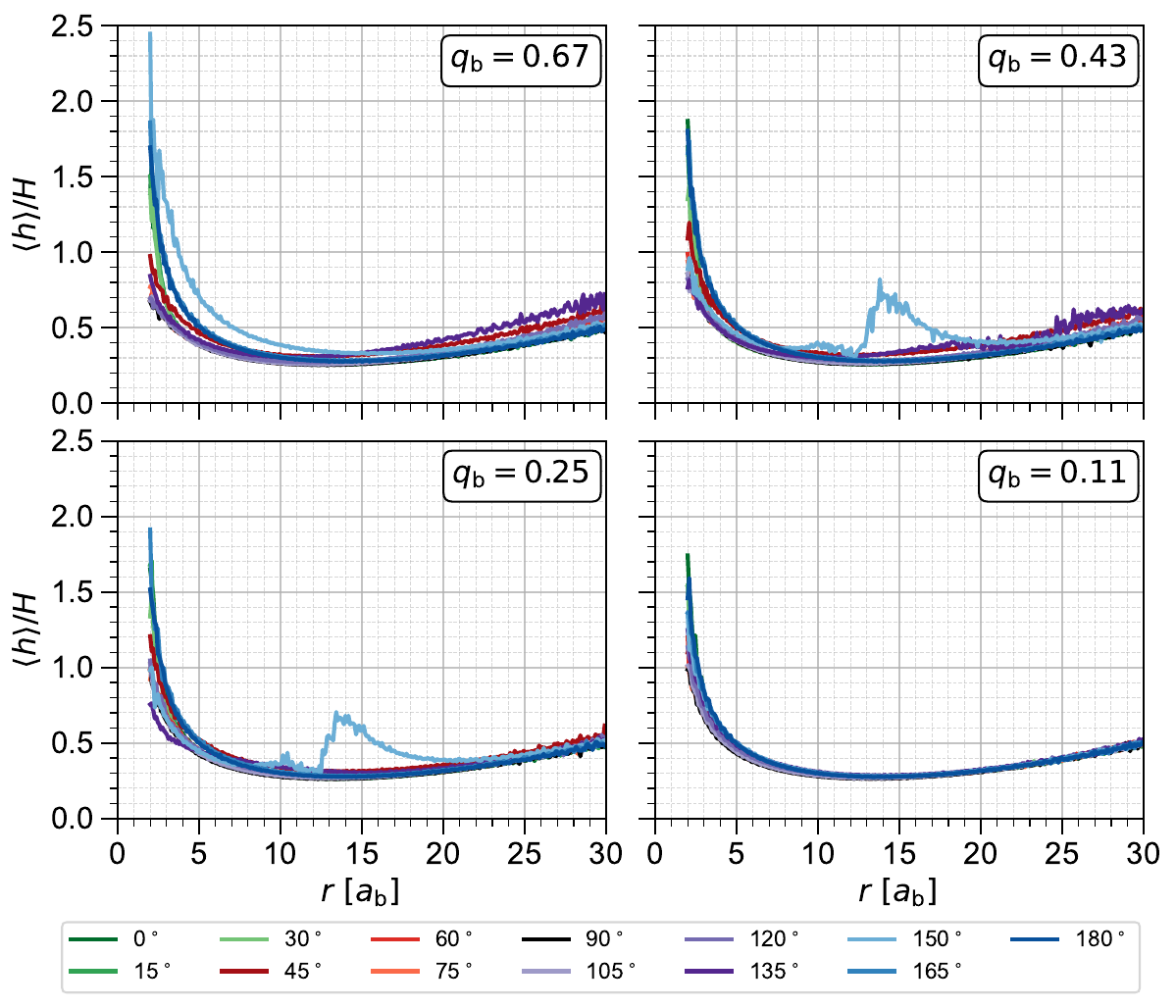}
		\caption{The average smoothing length per scale height $\langle h\rangle/H$, measured from radius $r = 2 a_\mathrm{b}$ to $30 a_\mathrm{b}$ at 5000 $P_\mathrm{b}$. $\langle h\rangle/H$ remains below 1.0 from $r = 4 a_\mathrm{b}$ to $30 a_\mathrm{b}$, indicating that the discs are well resolved. The peaks in panels for $q_\mathrm{b} = 0.43$ and $0.25$ correspond to the breaks.}
		\label{fig:resolution_check}
	\end{figure}

    Therefore, we examined $\langle h\rangle/H$ for all simulations  at the final simulation time. Figure~\ref{fig:resolution_check} shows the average smoothing length per scale height, $\langle h\rangle/H$, from $r = 2 a_\mathrm{b}$ to $30 a_\mathrm{b}$ for all simulations. Within $4 a_\mathrm{b}\le r \le 30 a_\mathrm{b}$, $\langle h\rangle/H$ remains below 0.8 in all cases, indicating that the circumbinary discs are well resolved. In cases with $i_0 = 150^\circ$ for $q_\mathrm{b}\in\{0.43,\ 0.25\}$, the peaks in $\langle h \rangle / H$ correspond to the locations of the disc breaks, where $\langle h \rangle / H$ increases substantially but remains below 0.8. Thus, studying the evolution of the circumbinary disc in our work is feasible.
    
    However, the discs become unresolved at smaller radii $r \le 4 a_\mathrm{b}$. A caveat here is that the accretion streams within the cavity and the individual circumstellar discs are not well resolved. Considering that the binary orbital parameters depend sensitively on these structures, accurate accretion rates and orbital evolution cannot be obtained from our simulations. Nevertheless, we can still analyse the accretion rate patterns of the binary.
    
    \section{Long-term accretion}
	\label{appendix:Accretion}
\begin{figure*}
		\centering
		\begin{subfigure}[b]{0.95\textwidth}
			\centering
			\includegraphics[width=\textwidth]{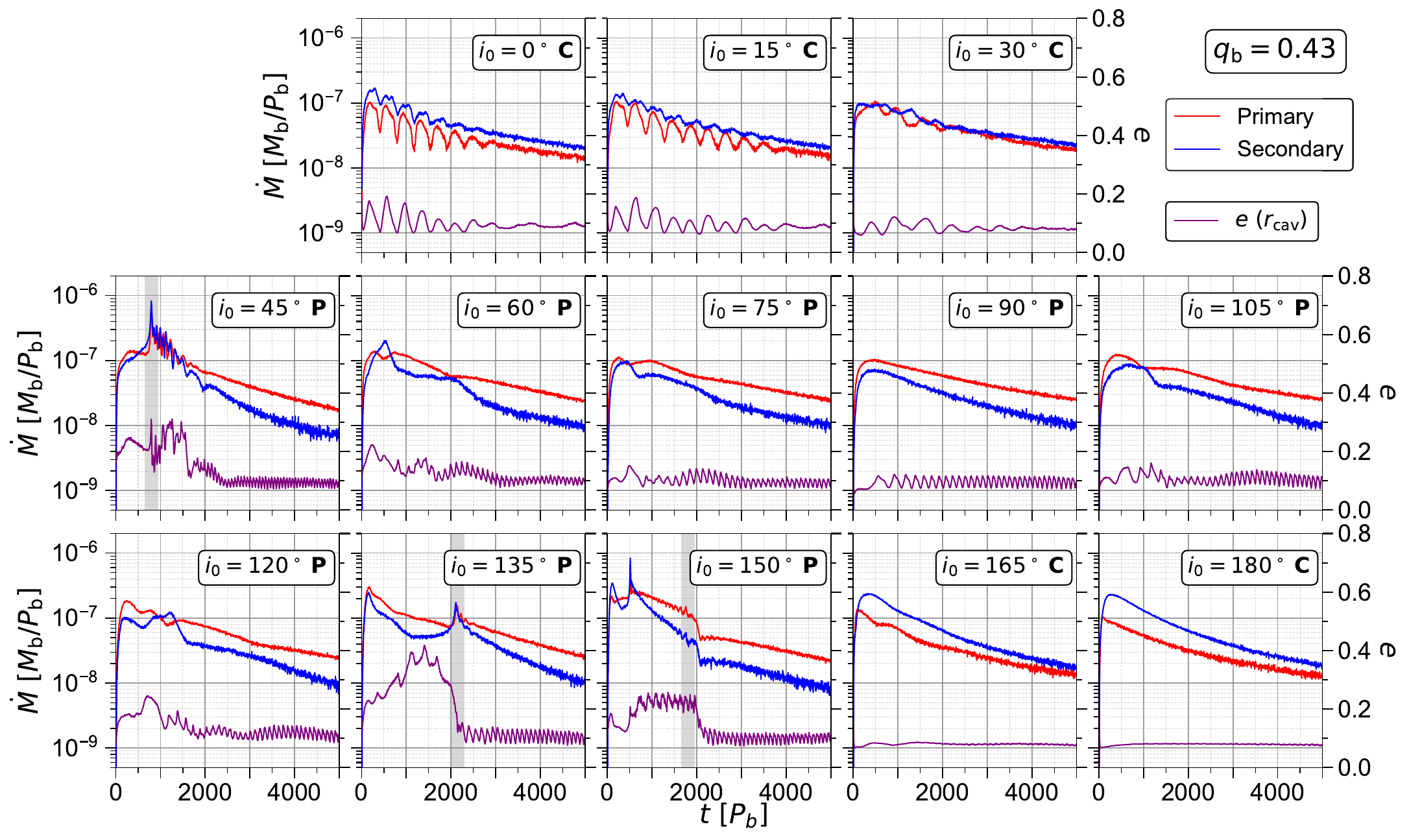}
			\label{fig:0p43_mdot}
		\end{subfigure}
		\caption{Same as Fig.~\ref{fig:mdot_0p67}, but for a binary mass ratio of $q_\mathrm{b}=0.43$.}
		\label{fig:mdot_0p43}
	\end{figure*}

\begin{figure*}
		\centering
		\begin{subfigure}[b]{0.95\textwidth}
			\centering
			\includegraphics[width=\textwidth]{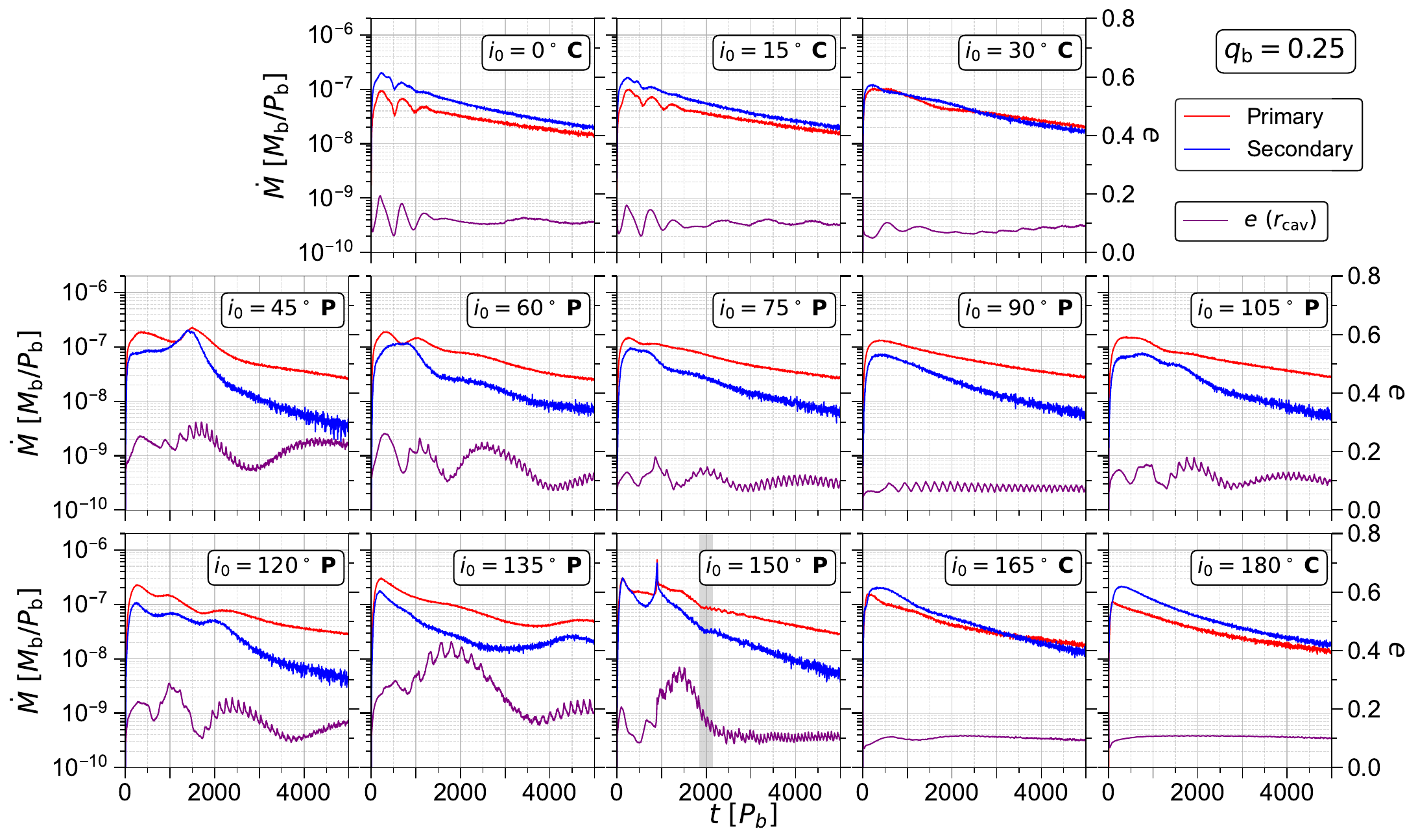}
			\label{fig:0p25_mdot}
		\end{subfigure}
		\caption{Same as Fig.~\ref{fig:mdot_0p67}, but for a binary mass ratio of $q_\mathrm{b}=0.25$.}
		\label{fig:mdot_0p25}
	\end{figure*}

\begin{figure*}
		\centering
		\begin{subfigure}[b]{0.95\textwidth}
			\centering
			\includegraphics[width=\textwidth]{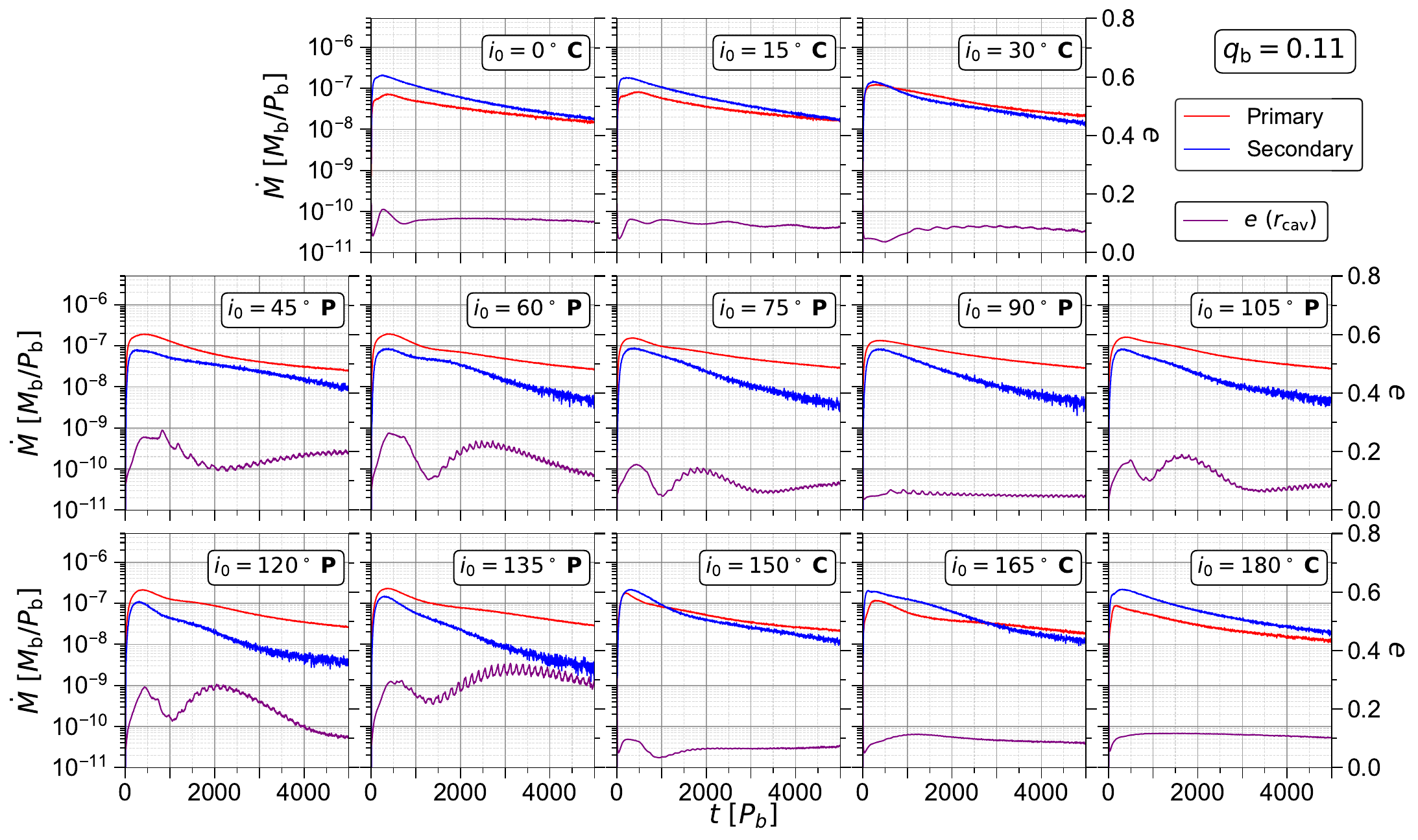}
			\label{fig:0p11_mdot}
		\end{subfigure}
		\caption{Same as Fig.~\ref{fig:mdot_0p67}, but for a binary mass ratio of $q_\mathrm{b}=0.11$.}
		\label{fig:mdot_0p11}
	\end{figure*}

\begin{figure*}
		\centering
		\begin{subfigure}[b]{0.95\textwidth}
			\centering
			\includegraphics[width=\textwidth]{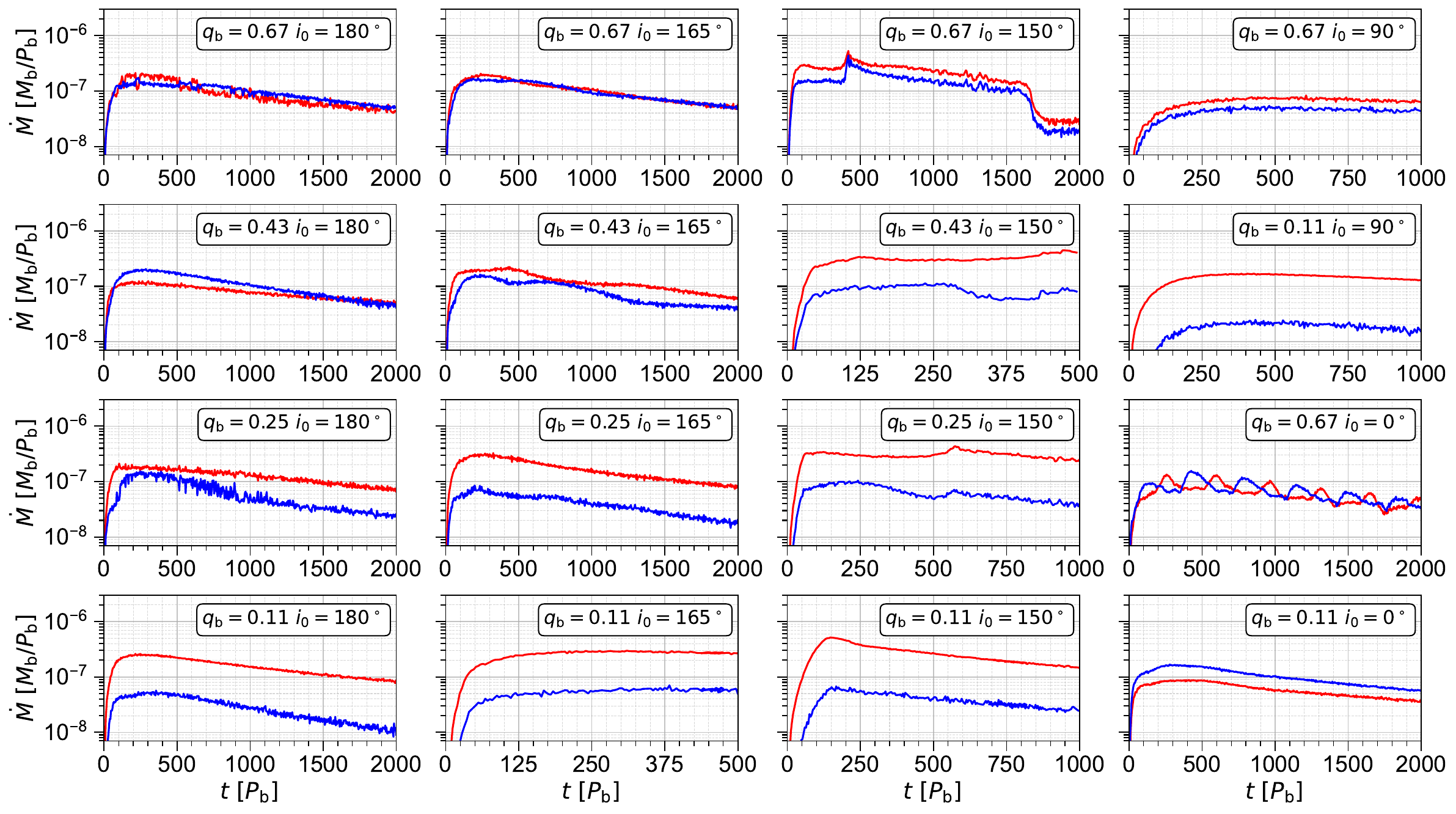}
		\end{subfigure}
		\caption{Binary accretion rates with an accretion radius of $r_\mathrm{acc}=0.025a_\mathrm{b}$. Red and blue curves represent the primary and secondary stars, respectively. Note that some runs were evolved only up to 1000$P_\mathrm{b}$ or 500$P_\mathrm{b}$.}
		\label{fig:compare_racc}
	\end{figure*}

       Figure~\ref{fig:mdot_0p43}, ~\ref{fig:mdot_0p25}, and ~\ref{fig:mdot_0p11} show the evolution of the long-term binary accretion rate ($\dot{M}$) and the inner disc eccentricity ($e$) at $r=4a_\mathrm{b}$ as a function of time, for binary mass ratios $q_\mathrm{b} = 0.43$, $0.25$, and $0.11$. Figure~\ref{fig:compare_racc} shows the binary accretion rates using $r_\mathrm{acc}=0.025a_\mathrm{b}$. Due to computational limitations, some runs were evolved only up to 1000$P_\mathrm{b}$ or 500$P_\mathrm{b}$.
	
	\bsp	
	\label{lastpage}
\end{document}